\documentclass[conference]{IEEEtran}
\usepackage{cite}
\usepackage{graphicx}
\usepackage{psfrag}
\usepackage{url}
\usepackage{amsmath}
\usepackage{array}
\usepackage{amssymb}
\usepackage{amsfonts}
\usepackage{graphicx}
\usepackage{epstopdf}
\usepackage{algorithm}
\usepackage{algpseudocode}
\newtheorem{proposition}{Proposition}
\newtheorem{lemma}{Lemma}

\newtheorem{definition}{Definition}
\newtheorem{theorem}{Theorem}

\newtheorem{example}{Example}

\usepackage{setspace}
\usepackage{amscd}
\usepackage{mathrsfs}
\usepackage{epsfig}
\usepackage{color}
\usepackage{textcomp}
\usepackage{multirow}
\usepackage{caption}
\usepackage{subcaption}
\usepackage{threeparttable}

\title{Physical Layer Network Coding for the Multiple Access Relay Channel}
\begin{document}

\author{
\authorblockN{Vijayvaradharaj T. Muralidharan and B. Sundar Rajan}
\authorblockA{Dept. of ECE, IISc, Bangalore 560012, India, Email:{$\lbrace$tmvijay, bsrajan$\rbrace$}@ece.iisc.ernet.in
}
}

\maketitle
\begin{abstract}
We consider the two user wireless Multiple Access Relay Channel (MARC), in which nodes $A$ and $B$ want to transmit messages to a destination node $D$ with the help of a relay node $R$. For the MARC, Wang and Giannakis proposed a Complex Field Network Coding (CFNC) scheme. As an alternative, we propose a scheme based on Physical layer Network Coding (PNC), which has so far been studied widely only in the context of two-way relaying. For the proposed PNC scheme, transmission takes place in two phases: (i) Phase 1 during which $A$ and $B$ simultaneously transmit and, $R$ and $D$ receive, (ii) Phase 2 during which $A$, $B$ and $R$ simultaneously transmit to $D$. At the end of Phase 1, $R$ decodes the messages $x_A$ of $A$ and $x_B$ of $B,$ and during Phase 2 transmits $f(x_A,x_B),$ where $f$ is many-to-one. Communication protocols in which the relay node decodes are prone to loss of diversity order, due to error propagation from the relay node. To counter this, we propose a novel decoder which takes into account the possibility of an error event at $R$, without having any knowledge about the links from $A$ to $R$ and $B$ to $R$. It is shown that if certain parameters are chosen properly and if the map $f$ satisfies a condition called \textit{exclusive law}, the proposed decoder offers the maximum diversity order of two. Also, it is shown that for a proper choice of the parameters, the proposed decoder admits fast decoding, with the same decoding complexity order as that of the CFNC scheme. Simulation results indicate that the proposed PNC scheme performs better than the CFNC scheme.
\end{abstract}

\section{Background and Preliminaries}
\begin{figure}[htbp]
\centering
\begin{subfigure}[]{1.7in}
\includegraphics[totalheight=1.5in,width=1.85in]{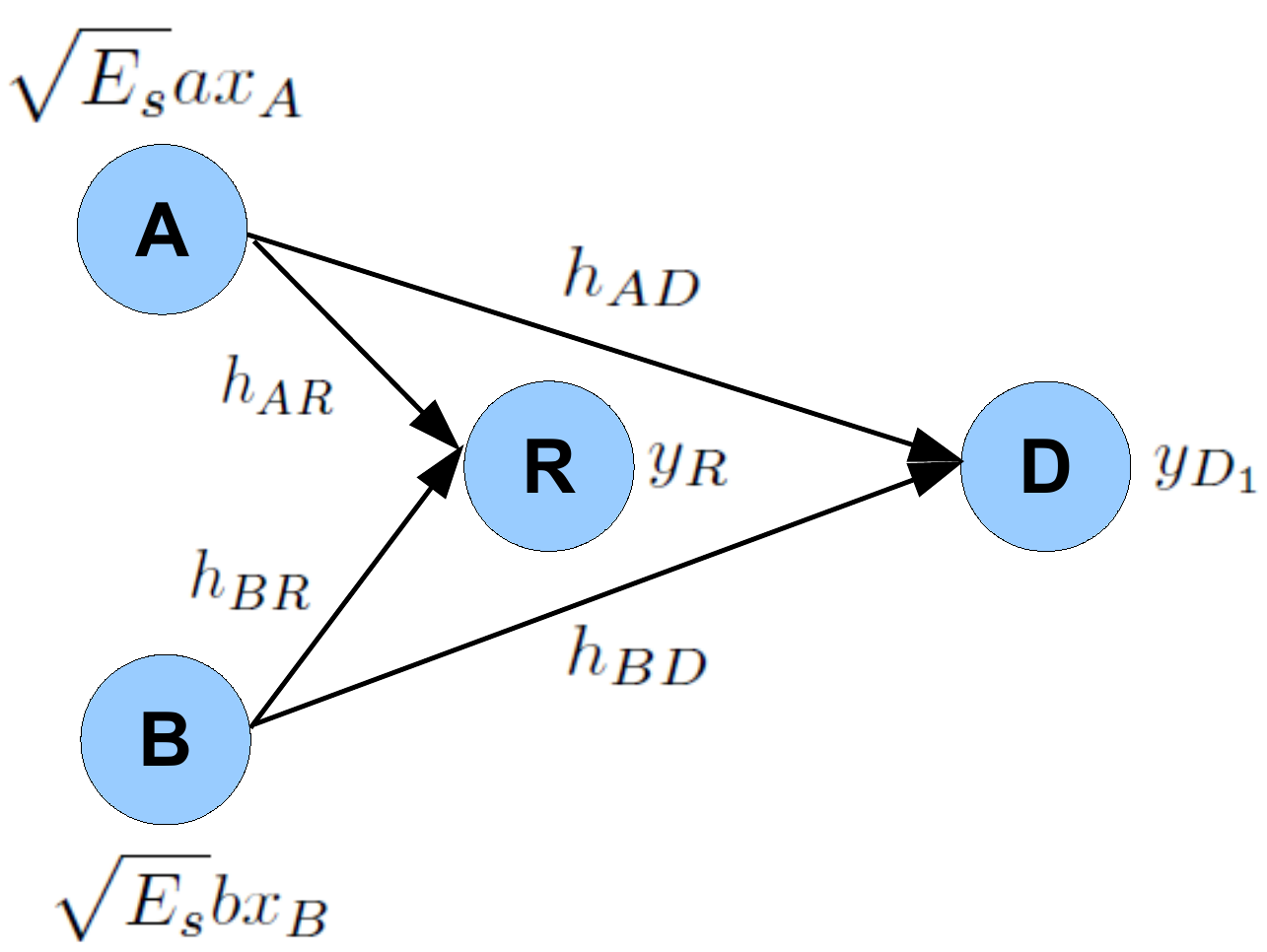}
\label{fig:phase1}
\caption{Phase 1}
\end{subfigure}
\begin{subfigure}[]{1.7in}
\includegraphics[totalheight=1.5in,width=1.85in]{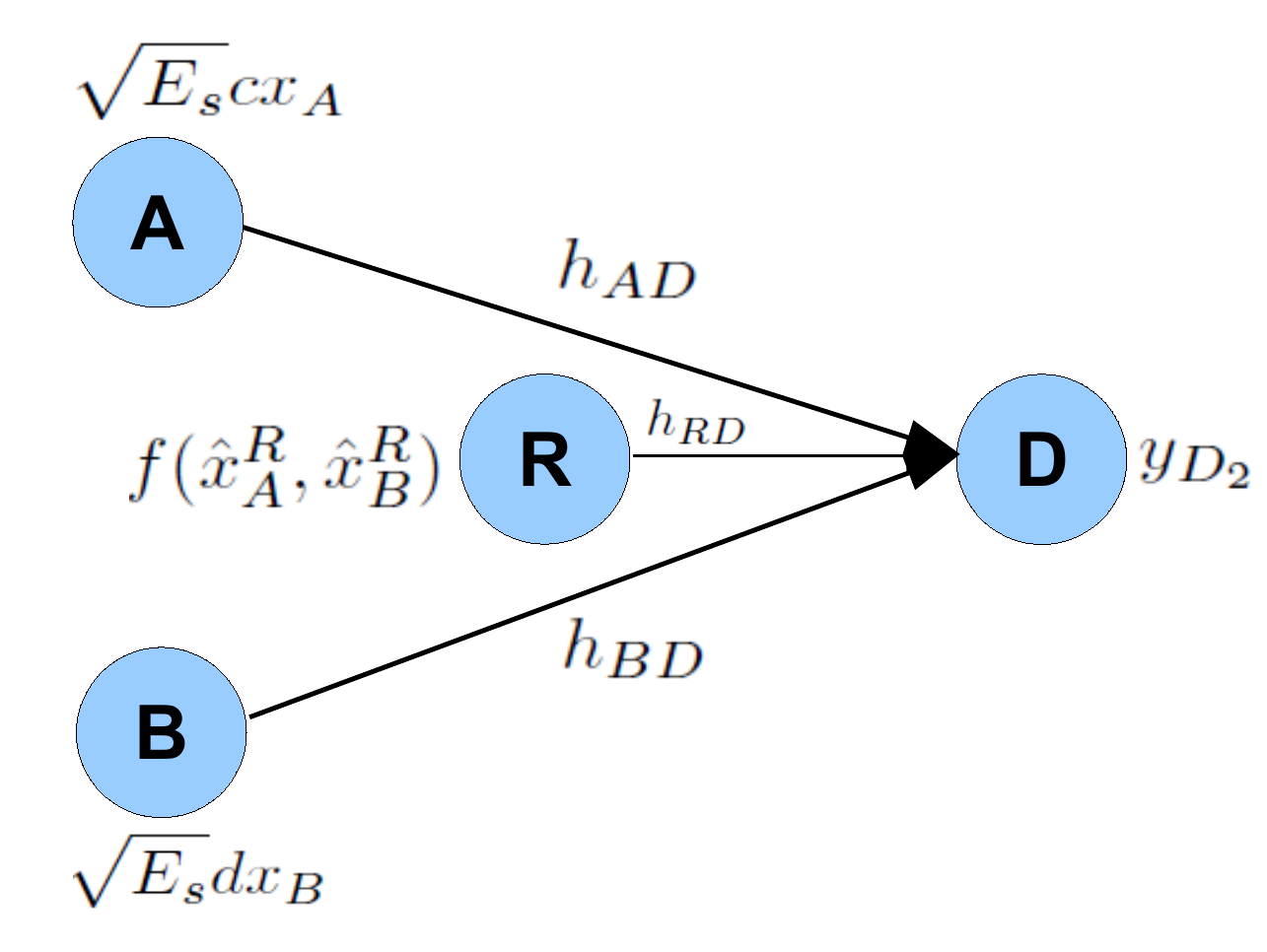}
\label{fig:phase2}
\caption{Phase 2}
\end{subfigure}
\caption{The two user Multiple Access Relay Channel}
\label{fig:MARC}
\end{figure}

We consider the two user Multiple Access Relay Channel (MARC) shown in Fig. \ref{fig:MARC}. Source nodes $A$ and $B$ want to transmit messages to the destination node $D$ with the help of the relay node $R.$ All the nodes are assumed to have half-duplex constraint, i.e., the nodes cannot transmit and receive simultaneously in the same frequency band. In addition to the presence of direct link, communication paths exist from the source nodes $A$ and $B$ to the destination node $D,$ via the relay node $R.$ As a result, in a two user MARC channel, a diversity order of two can be achieved, if the transmission scheme is chosen properly. 
\subsection{Background}
In a wireless network, due to the superposition nature of the wireless channel, signals interfere at the nodes. Avoiding this interference by making the nodes transmit in orthogonal time/frequency slots incurs a loss of spectral efficiency. The concept of physical layer network coding, in which the nodes are allowed to transmit simultaneously resulting in interference, was first introduced in \cite{ZhLiLa}. Physical layer Network Coding (PNC) has been shown to outperform traditional schemes which involve orthogonal transmissions \cite{ZhLiLa}--\cite{WiNaPfSp}. So far, most of the works on physical layer network coding have mainly focussed only on the two-way relay channel. In this paper, we propose a scheme based on PNC for the MARC. 

In a two-way relay channel, in order to ensure unique decodability at the end nodes, the network coding maps used at the relay node should satisfy a condition called the exclusive law \cite{APT1}. These maps satisfying the exclusive law form a mathematical structure called Latin Squares and the properties of Latin Squares have been used to obtain the network coding maps in a two-way relay channel \cite{NVR}--\cite{NR_Glcom}. An interesting connection between the proposed PNC scheme for the MARC and the two-way relay channel is that the network coding map used at $R$ needs to satisfy the exclusive law, for the proposed PNC scheme for the MARC to achieve a maximum diversity order two. 

Wireless relay networks, in which the relay nodes decode the messages are prone to loss of diversity order, due to the forwarding of erroneous messages. Various methods have been proposed in the literature to avoid this loss of diversity order. Cyclic Redundancy Check bits are used so that the nodes forward only those packets which are decoded correctly \cite{JaHeHuNo}. Some works assume the knowledge of all the instantaneous fade coefficients or error probabilities associated with the intermediate nodes at the destination node, with the decoder at the destination using this knowledge to ensure full diversity \cite{WaCaGiLa},\cite{JuKi}. Another method used widely is to use a scaling factor at the relay nodes which depends on the fade coefficients, with the scaling factor indicated to the destination using pilot symbols \cite{WaGiWa},\cite{WaGi}. The proposed scheme does not suffer from the disadvantages of any of the above methods, yet ensures maximum diversity order. This is achieved by means of an efficient choice of the transmission scheme and a novel decoder used at the destination $D.$

 A Complex Field Network Coding (CFNC) scheme for the MARC was proposed in \cite{WaGi}. 
 In this paper, as an alternative,  we propose a PNC scheme. As observed in \cite{WaGi}, when $R$ transmits a many-to-one function of $A$'s and $B$'s transmission during the relaying phase and minimum squared Euclidean distance decoder employed at $D$, a loss of diversity order results. In this paper, we show that for the proposed PNC scheme, making the source nodes also transmit during the relaying phase, combined with a novel decoder which is not minimum squared Euclidean distance decoder, ensures the maximum possible diversity order of two. Furthermore, if certain parameters are chosen properly, the proposed decoder for the PNC scheme can be implemented with a decoding complexity order same as that of the CFNC scheme.
  
 The following are the main advantages of the proposed PNC scheme over the CFNC scheme proposed in \cite{WaGi}: 
\begin{itemize}
\item 
 In the CFNC scheme, $R$ transmits a complex linear combination of $A$'s and $B$'s messages and the signal set used at $R$ during the relaying phase has $M^2$ points, where $M$ is the size of the signal set used at $A$ and $B$. In contrast, since the proposed PNC scheme uses a many-to-one map, the signal set used during the relaying phase has only $M$ points and hence the PNC scheme is expected to perform better than the CFNC scheme, which is confirmed by the simulation results. 
 \item
 In the CFNC scheme, $R$ uses a scaling factor which is a function of the fade coefficients of the $A$-$R$ and $B$-$R$ links, which needs to  be indicated to $D$ using pilot symbols. Since the proposed PNC scheme does not involve any such scaling factor, there is no need of such pilot symbols.
 \end{itemize}
\textbf{\textit{Notations}:}
   Throughout, vectors are denoted by bold lower case letters and matrices are denoted by bold capital letters. The set of complex numbers is denoted by $\mathbb{C}.$ $\mathcal{CN}(0, \sigma ^2)$ denotes a circularly symmetric complex Gaussian random variable with mean zero and variance $\sigma ^2$ and $\mathcal{N}(0, \sigma ^2)$ denotes a Gaussian random variable with variance $\sigma^2.$ For a matrix $\mathbf{A},$ $\mathbf{A^T}$ and $\mathbf{A^*}$ denotes its transpose and conjugate transpose respectively. For a matrix $\mathbf{A},$ $\mathrm{rank}(\mathbf{A})$ denotes its rank and  $\det(\mathbf{A})$ denotes its determinant. For a complex number $x,$ $x^*$ denotes its conjugate and $\vert x \vert$ denotes its absolute value. For a vector $\mathbf{v},$ ${\| \mathbf{v} \|}$ denotes its Euclidean norm. The total transmission energy of all the three nodes is assumed to be equal to $E_s$ and all the additive noises are assumed to have a variance equal to $1.$ By SNR, we denote the transmission energy ${E_s}.$ For a signal set $\mathcal{S},$ $\Delta \mathcal{S}$ denotes the difference signal set of $\mathcal{S},$ $\Delta\mathcal{S}=\lbrace x-x' \vert x, x'\in \mathcal{S}\rbrace.$ The all zero matrix of size $n \times n$ is denoted by $\mathbf{O_n}.$ $\mathbb{E}(X)$ denotes the expectation of $X.$
   
 \subsection{Signal Model}
Throughout, a quasi-static fading scenario is assumed with the channel state information available only at the receivers. 

Let $\mathcal{S}$ denote the signal set of unit energy used at $A$ and $B$, with $M=2^\lambda$ points, $\lambda$ being a positive integer. Assume that $A$ and $B$ want to transmit $\lambda$-bit binary tuples to $D$. Let $\mu: \mathbb{F}_2^\lambda \rightarrow \mathcal{S}$ denote the mapping from bits to complex symbols used at $A$ and $B$. 

For the proposed PNC scheme, transmission occurs in two phases: Phase 1 during which $A$ and $B$ simultaneously transmit and, $R$ and $D$ receive, followed by the Phase 2 during which $A$, $B$ and $R$ transmit to $D$.
\subsubsection*{Phase 1}
 Let $x_A= \mu(s_A)$, $x_B=\mu(s_B)$ $\in \mathcal{S}$ denote the complex symbols  $A$ and $B$ want to convey to $D$, where $s_A,s_B \in \mathbb{F}_2^\lambda$. During Phase 1, $A$ and $B$ transmit scaled versions of $x_A$ and $x_B$ respectively. The received signal at $R$ and $D$ during Phase 1 are respectively given by,
{
\begin{align}
\nonumber
&y_R=h_{AR} \sqrt{E_s} a x_A + h_{BR} \sqrt{E_s} b x_B +z_R, \text{and}\\
\label{eqn_yd1}
&y_{D_1}=h_{AD} \sqrt{E_s} a x_A + h_{BD} \sqrt{E_s} b x_B +z_{D_1},
\end{align}
}where $a,b \in \mathbb{C}$ are constants and the additive noises $z_R$ and $z_{D_2}$ are assumed to be $\mathcal{CN}(0,1).$ The fade coefficients are Rayleigh distributed, with $h_{AR} \sim \mathcal{CN}(0,\sigma_{AR}^2),$ $h_{BR} \sim \mathcal{CN}(0,\sigma_{BR}^2),$ $h_{AD} \sim \mathcal{CN}(0,\sigma_{AD}^2)$ and $h_{BD} \sim \mathcal{CN}(0,\sigma_{BD}^2).$

Let $(\hat{x}_A^R,\hat{x}_B^R) \in \mathcal{S}^2$ denote the Maximum Likelihood (ML) estimate of $({x}_A,{x}_B)$ at $R$ based on the received complex number $y_{R}$, i.e.,
{\small $$\displaystyle(\hat{x}_A^R,\hat{x}_B^R)=\arg\min_{({x}'_A,{x}'_B) \in \mathcal{S}^2} \vert y_R-h_{AR} \sqrt{E_s} a {x}'_A-h_{BR} \sqrt{E_s} b {x}'_B\vert.$$}
\subsubsection*{Phase 2}
During Phase 2, $A$ and $B$ transmit scaled versions of $x_A$ and $x_B$ respectively and $R$ transmits $x_R=f(\hat{x}^R_A,\hat{x}^R_B),$ where $f:\mathcal{S}^2 \rightarrow \mathcal{S}$ is a many-to-one map. The received signal at $D$ during Phase 2 is given by,  

{\vspace{-.3 cm}
\small
\begin{align}
\label{eqn_yd2}
y_{D_2}=h_{AD} \sqrt{E_s} c x_A + h_{BD} \sqrt{E_s} d x_B +h_{RD}\sqrt{E_s}x_R+z_{D_2}.
\end{align} 
}where $c,d \in \mathbb{C}$ are constants and the additive noise $z_{D_2}$ is assumed to be $\mathcal{CN}(0,1).$ The fade coefficient $h_{RD}$ is assumed to be $\mathcal{CN}(0,\sigma_{RD}^2).$

In order to ensure that the total transmission energy at the nodes $A$ and $B$ is equal to $E_s,$ the constants $a,b,c$ and $d$ are chosen such that $\vert a\vert^2+\vert c \vert ^2 =1$ and $\vert b\vert^2+\vert d\vert^2=1.$

From \eqref{eqn_yd1} and \eqref{eqn_yd2}, the received complex numbers at $D$ during the two phases can be written in vector form as, 
\begin{align}
\nonumber
\underbrace{\begin{bmatrix} y_{D_1} & y_{D_2} \end{bmatrix}}_{\mathbf{y_D}}&=  \sqrt{E_s}\underbrace{\begin{bmatrix} h_{AD} & h_{BD} & h_{RD} \end{bmatrix}}_{\mathbf{h}} \underbrace{\begin{bmatrix} a x_A & c x_A\\b x_B & d x_B \\ 0  & x_R\end{bmatrix}}_{\mathbf{C}(x_A,x_B,x_R)} \\
\label{eqn_vform}
& \hspace{4 cm}+\underbrace{\begin{bmatrix} z_{D_1} & z_{D_2} \end{bmatrix}}_{\mathbf{z_D}}.
\end{align}
 The matrix $\mathbf{C}(x_A,x_B,x_R)$ in \eqref{eqn_vform} is referred to as the codeword matrix. The restriction of $\mathbf{C}(x_A,x_B,x_R)$ to the first two rows, denoted by $\mathbf{C_r}(x_A,x_B)$ is referred as the restricted codeword matrix, i.e. $\mathbf{C_r}(x_A,x_B)={\begin{bmatrix} a x_A & c x_A\\b x_B & d x_B \end{bmatrix}}.$ The matrices  $\mathbf{C_r}(\Delta x_A,\Delta x_B)={\begin{bmatrix} a \Delta x_A & c \Delta x_A\\b \Delta x_B & d \Delta x_B \end{bmatrix}},$ where $\Delta x_A, \Delta x_B \in \Delta \mathcal{S}$ are referred to as the restricted codeword difference matrices. 
 
From \eqref{eqn_vform}, the vector $\mathbf{y_D}$ can also be written as, $$\mathbf{y_D}=\sqrt{E_s}x_A \mathbf{h}\mathbf{W_A}+\sqrt{E_s}x_B \mathbf{h} \mathbf{W_B} +\sqrt{E_s}x_R \mathbf{h} \mathbf{W_R}+\mathbf{z_D},$$ where the matrices $\mathbf{W_A}=\begin{bmatrix} a &  c \\ 0  & 0 \\0 & 0 \end{bmatrix},$ $\mathbf{W_B}=\begin{bmatrix} 0 &  0 \\ b  & d \\0 & 0 \end{bmatrix},$ and $\mathbf{W_R}=\begin{bmatrix} 0 &  0 \\ 0  & 0 \\0 & 1 \end{bmatrix}$ are referred to as the weight matrices at node $A$, $B$ and R respectively.

The contributions and organization of the paper are as follows: 
A novel decoder for the proposed PNC scheme is presented in Section II A.
In Section II B, it is shown that the decoder presented in Section II A achieves a maximum diversity of two if and only if the following two conditions are satisfied: (i) the map $f$ satisfies the so called \textit{exclusive law} and (ii) the constants $a, b,c$ and $d$ are such that the restricted codeword difference matrices have full rank for all non-zero values of $\Delta x_A$ and $\Delta x_B.$
In Section III, the condition under which the proposed decoder admits fast decoding is obtained. It is shown that when the weight matrices $\mathbf{W_A}$ and $\mathbf{W_R}$ (or $\mathbf{W_B}$ and $\mathbf{W_R}$) are Hurwitz-Radon orthogonal, the proposed decoder admits fast decoding, with the decoding complexity order same as that of the CFNC scheme proposed in \cite{WaGi}.
Simulation results which show that the proposed PNC scheme performs better than the CFNC scheme are presented in Section IV. 
\section{A Novel Decoder for the Proposed PNC Scheme and its Diversity Analysis}
In Section II A, a novel decoder for the proposed PNC scheme is presented. In Section II B, the condition under which the proposed decoder offers a maximum diversity order two is obtained.
\subsection{A Novel Decoder for the Proposed PNC Scheme}
Consider the case when $D$ uses the minimum squared Euclidean distance decoder given by, 

{\vspace{-.2 cm}
\scriptsize
\begin{align*}
\left(\hat{x}_A^D, \hat{x}_B^D\right)&= \arg \min_{\left(x_A,x_B\right) \in \mathcal{S}^2} \left\lbrace\vert y_{D_1}-h_{AD} \sqrt{E_s} a \: x_A- h_{BD} \sqrt{E_s} b \: x_B\vert^2\right.\\
&\hspace{-.5 cm}\left.+ \vert y_{D_2} -h_{AD} \sqrt{E_s} c \: x_A- h_{BD} \sqrt{E_s} d \: x_B- h_{RD} \sqrt{E_s} f(x_A,x_B) \vert ^2\right\rbrace.
\end{align*}
\vspace{-.2 cm}}

\noindent Since the above decoder does not consider the possibility of error events at the relay node, it does not offer maximum transmit diversity order two.

Alternatively, we propose a novel decoder which considers the possibility of error events at $R$, given by, 
\begin{align}
\nonumber
\left(\hat{x}_A^D, \hat{x}_B^D\right)&= \arg \min_{\left(x_A,x_B\right)\in \mathcal{S}^2}  \left \lbrace m_1\left(x_A,x_B\right), \right.\\
\label{eqn_decoder}
&\hspace{2.4 cm}\left.\log \left(SNR\right)+m_2\left(x_A,x_B\right)\right \rbrace,
\end{align}
where the metrics $m_1\left(x_A,x_B\right)$ and $m_2\left(x_A,x_B\right)$ are given in \eqref{metric_m1} and \eqref{metric_m2} respectively, at the top of the next page.

\begin{figure*}
\scriptsize
\begin{align}
\label{metric_m1}
&m_1\left(x_A,x_B\right)=\left\vert y_{D_1}-h_{AD} \sqrt{E_s} a \: x_A- h_{BD} \sqrt{E_s} b \: x_B\right\vert^2+ \left\vert y_{D_2} -h_{AD} \sqrt{E_s} c \: x_A- h_{BD} \sqrt{E_s} d \: x_B- h_{RD} \sqrt{E_s} f(x_A,x_B) \right\vert ^2,\\
\label{metric_m2}
&m_2\left(x_A,x_B\right)=\left\vert y_{D_1}-h_{AD} \sqrt{E_s} a \: x_A- h_{BD} \sqrt{E_s} b \: x_B\right\vert^2+\min_{x_R \neq f\left(x_A,x_B\right), x_R \in \mathcal{S}}\left \lbrace \left\vert y_{D_2} -h_{AD} \sqrt{E_s} c \: x_A- h_{BD} \sqrt{E_s} d \: x_B- h_{RD} \sqrt{E_s} x_R \right\vert ^2\right \rbrace.\\
\hline
\label{metric_m3}
&m_3\left(x_A,x_B\right)=\left\vert y_{D_1}-h_{AD} \sqrt{E_s} a \: x_A- h_{BD} \sqrt{E_s} b \: x_B\right\vert^2+\min_{x_R \in \mathcal{S}}\left \lbrace \left\vert y_{D_2} -h_{AD} \sqrt{E_s} c \: x_A- h_{BD} \sqrt{E_s} d \: x_B- h_{RD} \sqrt{E_s} x_R \right\vert ^2\right \rbrace.
\end{align}
\hrule
\end{figure*}
 
The idea behind the choice of this decoder is as follows: 
If the relay transmits the correct network-coded symbol, the optimal ML decoding metric at $D$ is given by $m_1(x_A,x_B).$
 The relay transmits a wrong network-coded symbol, independent of $(x_A,x_B),$ if the joint ML estimate at the relay $(\hat{x}_A^R, \hat{x}_B^R)$ is such  that $x_R=f(\hat{x}_A^R, \hat{x}_B^R) \neq  f(x_A,x_B).$ Under this condition, the optimal ML decision metric at $D$ is given by $m_2(x_A,x_B).$
The relay transmits a wrong network-coded symbol with a probability which is proportional to $\frac{1}{SNR}$ at high SNR. Hence to the metric $m_2(x_A,x_B),$ a correction factor $\log(SNR)$ is added and the minimum of $m_1(x_A,x_B)$ and $\log(SNR)+m_2(x_A,x_B)$ is taken to be the decoding metric at $D$.

The CFNC scheme proposed in \cite{WaGi} uses minimum squared Euclidean distance decoder, which has a decoding complexity of $\mathcal{O}(M^2).$ Since the decoder given in \eqref{eqn_decoder} involves minimization over three variables $x_A,x_B$ and $x_R,$ it appears as though the decoding complexity order is $\mathcal{O}(M^3).$ In Section III, it is shown that by properly choosing the constants $a,$ $b,$ $c$ and $d,$ the decoding complexity order can be reduced to $\mathcal{O}(M^2)$ which is the same as that of the CFNC scheme.

The diversity analysis of the decoder given in \eqref{eqn_decoder} is presented in the next subsection.
\subsection{Diversity Analysis of the Proposed Decoder}
The following theorem gives the condition under which the proposed decoder for the PNC scheme offers maximum diversity order two.
\begin{theorem}
For the proposed PNC scheme, the decoder given in \eqref{eqn_decoder} offers maximum diversity order two if and only if the following two conditions are satisfied:
\begin{enumerate}
\item
The map $f$ satisfies the condition called exclusive law given by,

{\vspace{-.2 cm}
\footnotesize
\begin{equation}
 \left.\begin{aligned}
f(x_A,x_B) \neq f(x'_A,x_B), \; \mathrm{for} \;x_A \neq x'_A, \; \forall \;x_B \in  \mathcal{S},\\
\label{ex_law}
f(x_A,x_B) \neq f(x_A,x'_B), \; \mathrm{for} \;x_B \neq x'_B, \; \forall \;x_A \in \mathcal{S}.
       \end{aligned}
 \right\}
\end{equation}}
\item
 The restricted codeword difference matrices $\mathbf{C_r}(\Delta x_A, \Delta x_B)$ have full rank, $ \forall \Delta x_A \neq 0,$ $\Delta x_B \neq 0$ and $\Delta x_A, \Delta x_B  \in \Delta \mathcal{S}.$
\end{enumerate} 
\begin{proof}
See Appendix.
\end{proof}   
\end{theorem}

Note that condition 2) does not demand full rank for all the restricted codeword difference matrices. In fact, whatever may be the choice of $a,b,c$ and $d,$ it is impossible to obtain full rank for the restricted codeword difference matrices of the form $\mathbf{C_r}(\Delta x_A,0)$ (and also $\mathbf{C_r}(0,\Delta x_B)$), since both the entries of the second row of $\mathbf{C_r}(\Delta x_A,0)$ are zeros. It suffices to ensure full rank for only those restricted code word difference matrices for which both $\Delta x_A$ and $\Delta x_B$ are non-zeros.

It is easy to find a map $f$ satisfying the exclusive law, since all maps satisfying the exclusive law form Latin Squares \cite{NVR}. 
\begin{definition}{\cite{Rod}}
A Latin Square L of order $M$ on the symbols from the set $\mathbb{Z}_t=\{0,1, \cdots ,t-1\}$ is an ${M}$ $\times$ ${M}$ array, in which each cell contains one symbol and each symbol occurs at most once in each row and column. 
\end{definition}
Two simple examples of Latin Squares are the Modulo-M Latin square and the Bit-wise XOR Latin Square. In the Modulo-$M$ Latin Square, a cell in the $M$ $\times$ $M$ is filled with the modulo-$M$ sum of the row index and column index. In the bit-wise XOR Latin Square, a cell in the $M$ $\times$ $M$ array is filled the bit-wise exclusive OR of the row index and column index represented in binary, after binary-to-decimal conversion.  For $M=4,$ the Modulo-4 Latin Square and the Bit-wise XOR Latin Square are as shown in Fig. \ref{Latin_Example}.
\begin{figure}[h]
\centering
\begin{subfigure}[t]{1in}
\begin{tabular}{c|c|c|c|c|}
\hline & 0 & 1 & 2 & 3\\
\hline 0 &  0&1  & 2 &3 \\ 
\hline 1 &1  &2 & 3 &0 \\ 
\hline 2 &2  &3  &0  &1 \\ 
\hline 3 & 3 &0  &1&2\\ 
\hline 
\end{tabular}
\caption{Modulo-4 Latin Square}
\end{subfigure}
\quad \quad \quad \quad 
\begin{subfigure}[t]{1in}
\begin{tabular}{c|c|c|c|c|}
\hline & 0 & 1 & 2 & 3\\
\hline 0 &  0&1  & 2 &3 \\ 
\hline 1 &1  &0 & 3 &2 \\ 
\hline 2 &2  &3  &0  &1 \\ 
\hline 3 & 3 &2  &  1&0\\ 
\hline 
\end{tabular}
\caption{Bit-wise XOR Latin Square}
\end{subfigure}
\caption[]{Examples of Latin Squares of order 4}
\label{Latin_Example}
\end{figure} 

In the following example, a choice of $a,$ $b,$ $c$ and $d$ is provided, which ensures that the restricted codeword difference matrices have full rank for all non-zero values of $\Delta x_A$ and $\Delta x_B.$
\begin{example}
Choosing $a=1,$ $b=\frac{1}{\sqrt{2}},$ $c=0$ and $d=\frac{1}{\sqrt{2}},$ the restricted codeword difference matrices are of the form $\mathbf{C_r}(\Delta x_A, \Delta x_B)= \begin{bmatrix} \Delta x_A & 0 \\ \frac{1}{\sqrt{2}} \Delta x_B & \frac{1}{\sqrt{2}} \Delta x_B \end{bmatrix}.$ $\mathbf{C_r}(\Delta x_A, \Delta x_B)$ is full rank for all $\Delta x_A,\Delta x_B \neq 0$ since $det\left(\mathbf{C_r}(\Delta x_A, \Delta x_B)\right)=\frac{1}{\sqrt{2}}\Delta x_A \Delta x_B \neq 0.$
\end{example}

A sufficient condition under which the decoder given in \eqref{eqn_decoder} admits fast decoding is obtained in the next section. 
\section{A Fast Decoding Algorithm for the Proposed Decoder}
In this section, it is shown that by properly choosing the constants $a,b,c$ and $d,$ the decoder given in \eqref{eqn_decoder} can be implemented efficiently by means of a fast decoding algorithm. 

Before presenting the algorithm, we introduce some notations.

The points in the signal set $\mathcal{S}$ are denoted by $s_i, 1 \leq i \leq M.$ 

From \eqref{eqn_vform}, the vector $\mathbf{y_D}^T$ can be written as,
\begin{align*}
\mathbf{y_D}^T=\underbrace{\begin{bmatrix} a h_{AD} & b h_{BD} & 0 \\c h_{AD} & d h_{BD}& h_{RD} \end{bmatrix}}_{\mathbf{H_{eq}}} \underbrace{\begin{bmatrix} x_A \\ x_B \\ x_R \end{bmatrix}}_{\mathbf{x}} \sqrt{E_s}+\mathbf{z_D}^T.
\end{align*}
The matrix $\mathbf{H_{eq}}$ can be decomposed using $\mathbf{QR}$ decomposition as $\mathbf{H_{eq}}=\mathbf{Q}\mathbf{R},$ where $\mathbf{Q}$ is a $2 \times 2$ unitary matrix and $\mathbf{R}=[\mathbf{R_1} \; \mathbf{r_2}]$ is a $2 \times 3$ matrix, with $\mathbf{R_1}$ being upper-triangular of size $2 \times 2$ and $\mathbf{r_2}$ being a column vector of length 2. Let $r_{ij}$ denote the $(i,j)^{th}$ entry of $\mathbf{R}.$

Define $\mathbf{\tilde{y}_D}=\mathbf{Q}^T \mathbf{y_D}^T=[\tilde{y}_{D_1} \: \tilde{y}_{D_2}]^T.$

Also, let
\begin{align*}
&\phi_1(x_A,x_B)=\vert \tilde{y}_{D_1}-r_{11}x_A \sqrt{E_s}-r_{12}x_B \sqrt{E_s}\vert^2,\\
&\phi_2(x_A,x_B)=\vert \tilde{y}_{D_2} -r_{22}x_B \sqrt{E_s}- r_{23} f(x_A,x_B) \sqrt{E_s} \vert ^2 \text{and}\\
&\phi_3(x_B,x_R)=\vert \tilde{y}_{D_2}-r_{22}x_B \sqrt{E_s}-r_{23}x_R \sqrt{E_s}\vert^2.
\end{align*}
The following proposition gives a sufficient condition under which {\bf Algorithm 1} implements the decoder given in \eqref{eqn_decoder}.
\begin{algorithm}
{\small
\begin{algorithmic}[1]
\For {$i=1$ to $M$}
    \State $x_B\gets s_i$
    \State \textbf{Find} $\hat{x}_{A}^1=\displaystyle{\arg\min_{x_A \in \mathcal{S}}\lbrace \phi_1(x_A,x_B) + \phi_2(x_A,x_B)\rbrace}$
     \State \textbf{Find} $\hat{x}_{A}^2=\displaystyle{\arg\min_{x_A \in \mathcal{S}}\lbrace \phi_1(x_A,x_B)\rbrace}$ 
     \State \textbf{Find} $\hat{x}_R=\displaystyle{\arg\min_{x_R \in \mathcal{S}}\lbrace \phi_3(x_B,x_R)\rbrace}$ 
\If {$\phi_1(\hat{x}_A^1,x_B)+\phi_2(\hat{x}_A^1,{x}_B)<$ $\phi_1(\hat{x}_A^2,x_B)+\phi_3(x_B,\hat{x}_R)$ $\hspace{6.6 cm}$ $~~~~~~~~~~~~~~~~~~~~~~~~~~~~~~~~~~~~~~~~~~~~~~~~~~~~~~~~~~~~~~~~~~+\log(SNR)$ $~~~~$}{\\$~~~~~~~m(x_B)=\phi_1(\hat{x}_A^1,x_B)+\phi_2(\hat{x}_A^1,{x}_B)$ \\$~~~~~~~\hat{x}_A(x_B)=\hat{x}_A^1$}
\Else \\
{$~~~~~~~m(x_B)=\phi_1(\hat{x}_A^2,x_B)+\phi_3(x_B,\hat{x}_R)+\log(SNR)$\\$~~~~~~~\hat{x}_A(x_B)=\hat{x}_A^2$}
\EndIf    
\EndFor\\
\textbf{Find} ${\displaystyle\hat{x}_B=\arg \min_{x_B \in \mathcal{S}} m(x_B)}$\\
$(\hat{x}_A^D,\hat{x}_B^D)=(\hat{x}_A(\hat{x}_B),\hat{x}_B)$
\end{algorithmic}}
\caption{Decoding Algorithm used at $D$}
\end{algorithm}
\begin{proposition}
{\bf Algorithm 1} implements the decoder in \eqref{eqn_decoder}, if the constants $a,b,c$ and $d$ are such that the weight matrices $\mathbf{W_A}$ and $\mathbf{W_R}$ are Hurwitz-Radon (H-R) orthogonal, i.e.,
$\mathbf{W_A} \mathbf{W_R}^{*}+\mathbf{W_R} \mathbf{W_A}^{*}=\mathbf{O_3}.$ \footnote{A algorithm exactly similar to {\bf Algorithm 1} can be used with the roles of $A$ and $B$ interchanged, if $\mathbf{W_B}$ and $\mathbf{W_R}$ are H-R orthogonal.}
\begin{proof}
The decoding metric of the decoder given in \eqref{eqn_decoder} can be written as,

{\vspace{-.3 cm}
\footnotesize
\begin{align}
\nonumber
&\min_{\left(x_A,x_B\right)} \left\lbrace m_1\left(x_A,x_B\right) ,\log\left(SNR\right)+ m_2\left(x_A,x_B\right) \right\rbrace\\
\nonumber
&\hspace{2 cm}=\min_{\left(x_A,x_B\right)} \lbrace m_1\left(x_A,x_B\right) ,\log\left(SNR\right)+ m_1\left(x_A,x_B\right),\\
\nonumber
&\hspace{5 cm}\log\left(SNR\right)+ m_2\left(x_A,x_B\right)\rbrace.\\
\label{decoding_metric}
&\hspace{1 cm}=\min_{\left(x_A,x_B\right)} \lbrace m_1\left(x_A,x_B\right) ,\log\left(SNR\right)+ m_3\left(x_A,x_B\right)\rbrace.
\end{align}
}where the metric $m_3(x_A,x_B)$ is given in \eqref{metric_m3}, at the top of this page.
%
We have,
\begin{align}
\nonumber
m_3(x_A,x_B)&= \min_{x_R \in \mathcal{S}} \lbrace \|\mathbf{y_{D}}^T-\mathbf{H_{eq}}\mathbf{x}\sqrt{E_s}\|^2\rbrace\\
\label{metric_m3_second}
&=\min_{x_R \in \mathcal{S}} \lbrace \|\mathbf{\tilde{y}_D}-\mathbf{R}\mathbf{x}\sqrt{E_s}\|^2\rbrace.
\end{align}
Since $\mathbf{R_1}$ is upper triangular, $r_{21}=0.$ Also, the entry $r_{13}=0,$ since $\mathbf{W_A}$ and $\mathbf{W_R}$ are H-R orthogonal (follows from Theorem 2, \cite{SrRa}). Hence, from \eqref{metric_m3_second}, it follows that,
\begin{align}
\nonumber
m_3(x_A,x_B)&=  \vert \tilde{y}_{D_1}-r_{11}x_A \sqrt{E_s}-r_{12}x_B \sqrt{E_s}\vert^2\\
&\hspace{-1 cm}+ \min_{x_R \in \mathcal{S}}\left \lbrace\vert \tilde{y}_{D_2}-r_{22}x_B \sqrt{E_s}-r_{23}x_R \sqrt{E_s}\vert^2\right\rbrace.
\end{align}
Hence, $$\displaystyle{\min_{x_A \in \mathcal{S}} m_3(x_A,x_B)= \min_{x_A \in \mathcal{S}}\phi_1(x_A,x_B)+\min_{x_R \in \mathcal{S}}\phi_3(x_B,x_R)}.$$

From \eqref{decoding_metric}, the decoding metric can be written as,
\begin{align*}
&\min_{x_B \in \mathcal{S}}\left\lbrace \min_{x_A \in \mathcal{S}}\left(\phi_1\left(x_A,x_B\right)+\phi_2\left(x_A,x_B\right)\right)\right., \\
&\left.\hspace{1 cm}\min_{x_A \in \mathcal{S}}\phi_1\left(x_A,x_B\right)+\min_{x_R \in \mathcal{S}}\phi_3\left(x_B,x_R\right)+\log\left(SNR\right)\right\rbrace
\end{align*}
\end{proof}
\end{proposition}
In {\bf Algorithm 1}, inside the {\bf for} loop, $x_B$ is fixed and the operations in lines 3, 4 and 5 entail a complexity order $\mathcal{O}(M).$ The operations from line 6 to line 12 involve constant complexity, independent of $M.$ Hence the complexity order for executing the {\bf for} loop from line 1 to line 13 is $\mathcal{O}(M^2).$ The operation in line 14 involves a complexity order $\mathcal{O}(M).$ Hence the overall decoding complexity order of {\bf Algorithm 1} is $\mathcal{O}(M^2),$ which is the same as that of the CFNC scheme proposed in \cite{WaGi}. 
\begin{example}
For the case when $a=1,$ $b=\frac{1}{\sqrt{2}},$ $c=0$ and $d=\frac{1}{\sqrt{2}},$ the weight matrices are given by, $\mathbf{W_A}=\begin{bmatrix}1 & 0\\0 & 0\\0 & 0\end{bmatrix},$ $\mathbf{W_B}=\begin{bmatrix}0 & 0\\\frac{1}{\sqrt{2}} & \frac{1}{\sqrt{2}}\\0 & 0\end{bmatrix}$ and $\mathbf{W_R}=\begin{bmatrix}0 & 0\\0 & 0\\0 & 1\end{bmatrix}.$ It can be verified that the matrices $\mathbf{W_A}$ and $\mathbf{W_R}$ are H-R orthogonal, i.e., $\mathbf{W_A} \mathbf{W_R}^{*}+\mathbf{W_R} \mathbf{W_A}^{*}=\mathbf{O_3}.$ Hence, for this case, {\bf Algorithm 1} can be used to implement the decoder given in \eqref{eqn_decoder}.
\end{example}
\section{Simulation Results}
Simulation results presented in this section compare the performance of the proposed PNC scheme with the CFNC scheme proposed in \cite{WaGi}. In all the simulation results presented, the values of the constants are chosen to be $a=1,$	$b=\frac{1}{\sqrt{2}},$ $c=0$ and $d=\frac{1}{\sqrt{2}}$ and 4-PSK signal set is used at the nodes. The Modulo-4 Latin Square is used at the relay node. 

\begin{figure}[htbp]
\centering
\vspace{-.75 cm}
\includegraphics[totalheight=2.8in,width=3.85in]{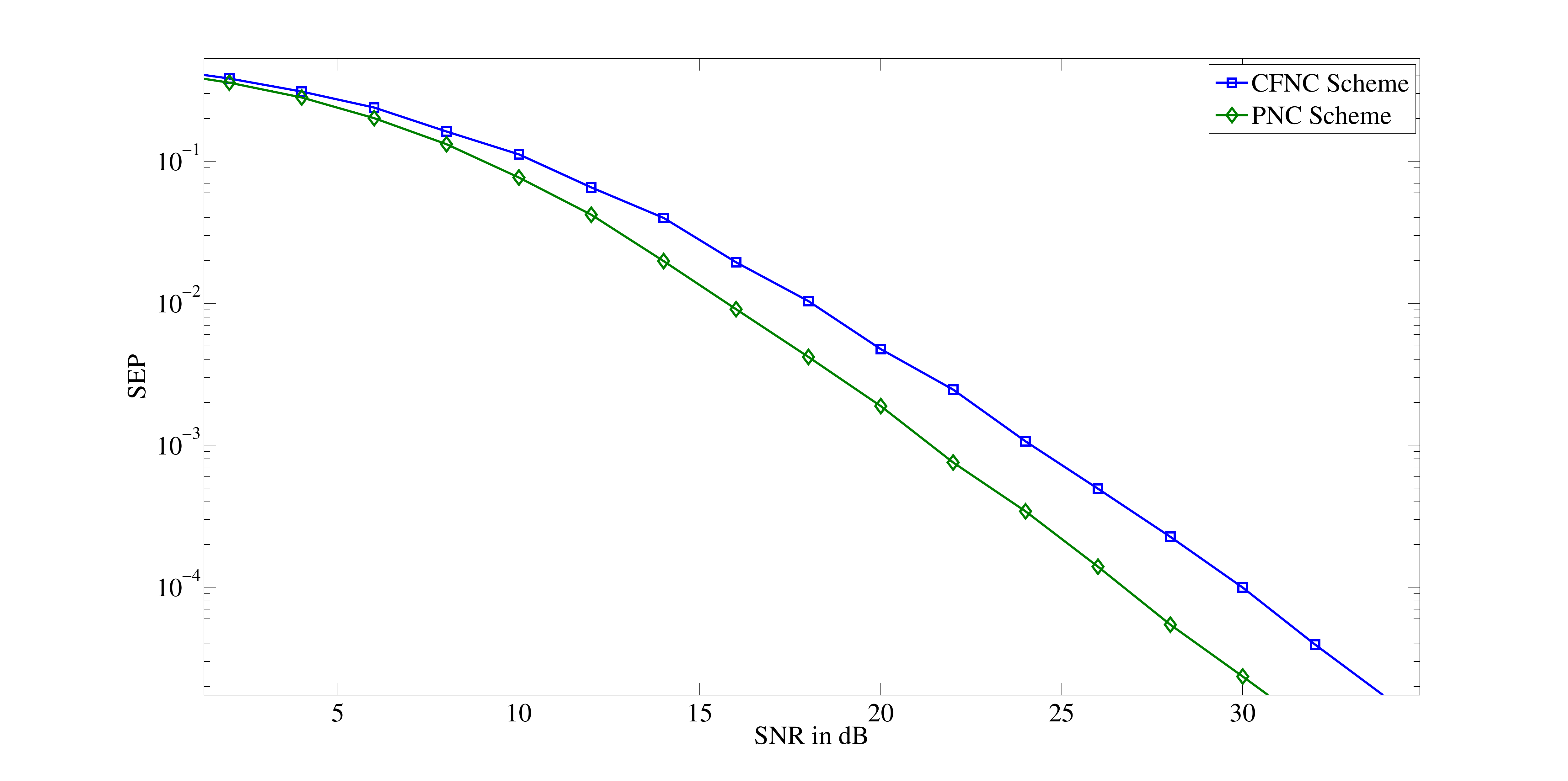}
\vspace{-.75 cm}
\caption{SNR vs SEP plots for the PNC and CFNC schemes for 4-PSK signal set for $\sigma_{AR}^2=\sigma_{BR}^2=\sigma_{AD}^2=\sigma_{BD}^2=\sigma_{RD}^2=0$ dB.}	
\label{fig:plots_4psk_equal}
\end{figure}

\begin{figure}[htbp]
\centering
\vspace{-.75 cm}
\includegraphics[totalheight=2.8in,width=3.85in]{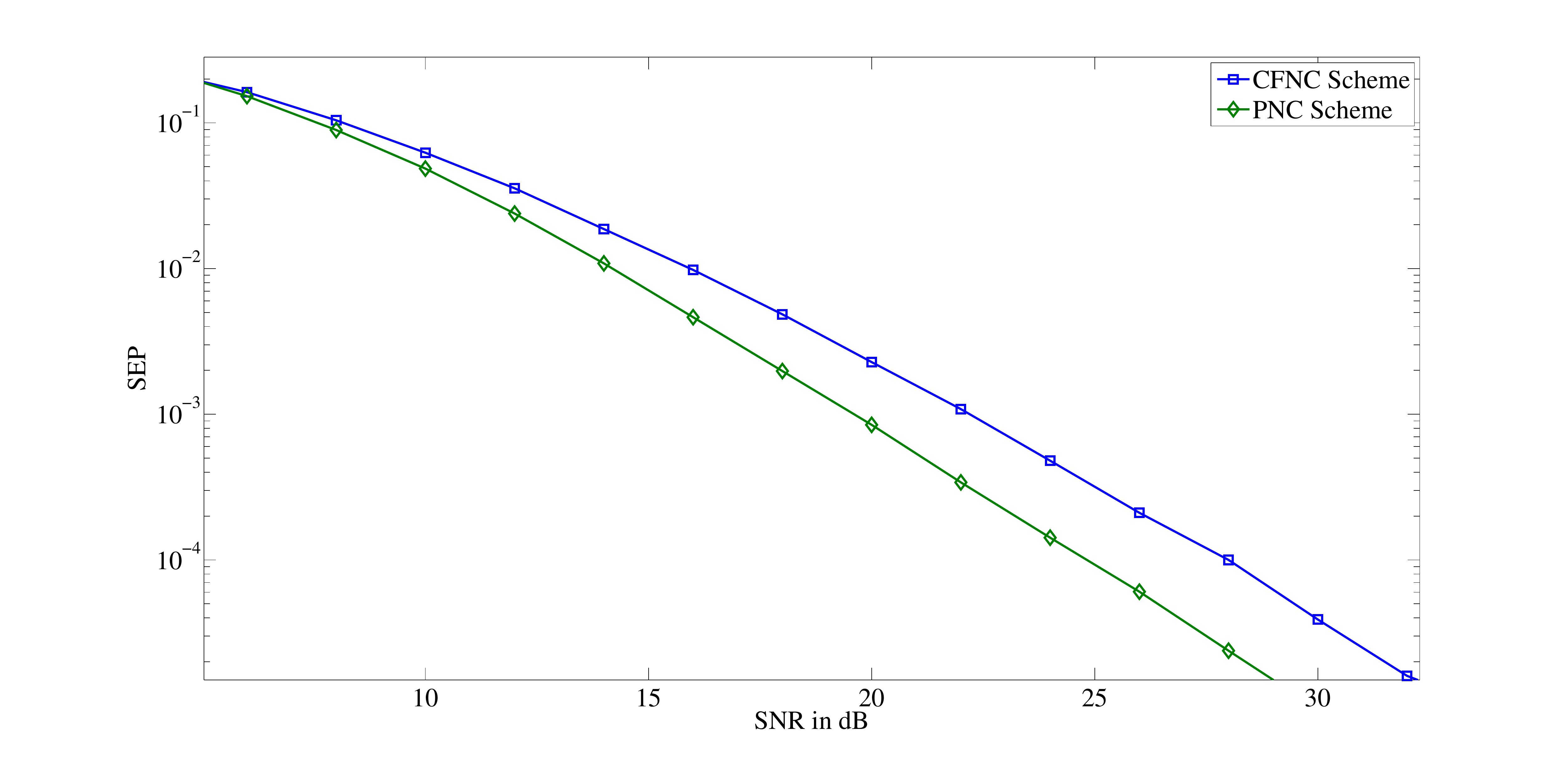}
\vspace{-.75 cm}
\caption{SNR vs SEP plots for the PNC and CFNC schemes for 4-PSK signal set for $\sigma_{AR}^2=\sigma_{BR}^2=10$ dB, $\sigma_{AD}^2=\sigma_{BD}^2=\sigma_{RD}^2=0$ dB.}	
\label{fig:plots_4psk_sr_strong}
\vspace{-.5 cm}	
\end{figure}

\begin{figure}[htbp]
\centering
\vspace{-.75 cm}
\includegraphics[totalheight=2.8in,width=3.85in]{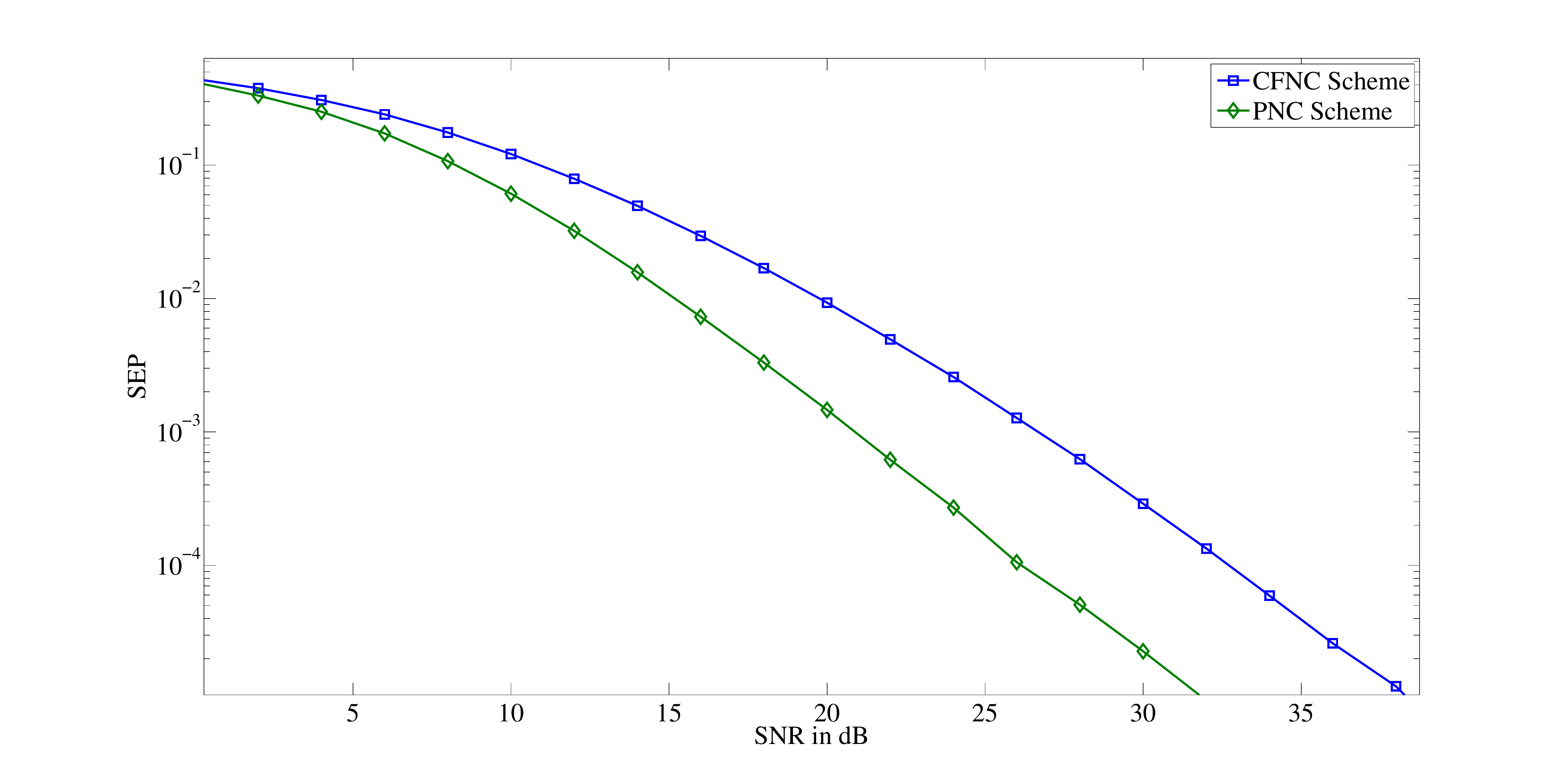}
\vspace{-.75 cm}
\caption{SNR vs SEP plots for the PNC and CFNC schemes for 4-PSK signal set for $\sigma_{AR}^2=\sigma_{BR}^2=\sigma_{AD}^2=\sigma_{BD}^2=0$ dB, $\sigma_{RD}^2=10$ dB.}	
\label{fig:plots_4psk_rd_strong}
\vspace{-.5 cm}	
\end{figure}

Fig. \ref{fig:plots_4psk_equal} shows the SNR Vs. Symbol Error Probability (SEP) plots for the case when the variances of all the fading links are 0 dB. It can be seen from Fig. \ref{fig:plots_4psk_equal} that the PNC scheme performs better than the CFNC scheme and offers a gain of nearly 3.3 dB at high SNR. Fig. \ref{fig:plots_4psk_sr_strong} shows a similar plot for the case when the links from $A$-$R$ and $B$-$R$ are stronger than the other links, i.e., $\sigma_{AR}^2=\sigma_{BR}^2=10$ dB, $\sigma_{AD}^2=\sigma_{BD}^2=\sigma_{RD}^2=0$ dB. It can be seen from Fig. \ref{fig:plots_4psk_sr_strong} that for this case, the PNC scheme offers a gain of nearly 3 dB at high SNR. Fig. \ref{fig:plots_4psk_rd_strong} shows the plots for the case when the $R$-$D$ link is stronger than all other links, i.e, $\sigma_{AR}^2=\sigma_{BR}^2=\sigma_{AD}^2=\sigma_{BD}^2=0$ dB, $\sigma_{RD}^2=10$ dB. For this case, the PNC scheme offers a gain of about 6.5 dB at high SNR. Also, it can be verified from the plots that the diversity order for the proposed PNC scheme is two.
\section{Discussion}
A scheme based on physical layer network coding was proposed for the Multiple Access Relay Channel (MARC). A novel decoder was proposed for the PNC scheme. The conditions which the network coding map $f$ and the constants $a,b,c$ and $d$ should satisfy, for the proposed decoder to offer a maximum diversity order two were obtained. It was shown that if the constants $a,b,c$ and $d$ are chosen properly, the proposed decoder can be implemented efficiently by a fast decoding algorithm. Simulation results presented showed that the proposed decoder performs better than the CFNC scheme proposed in \cite{WaGi}. The problem of optimizing the choice of the constants $a,$ $b,$ $c$ and $d$ to minimize the error probability in addition to ensuring maximum diversity remains open. Extending the scheme to a Multiple Access Relay network with more than two source nodes and multiple relay nodes is a possible direction for future work.

\section*{APPENDIX - Proof of Theorem 1}
Let $H=(h_{AR},h_{BR},h_{RD},h_{AD},h_{BD})$ denote a particular realization of the fade coefficients. Throughout the proof, the subscript $H$ in a probability expression indicates conditioning on the fade coefficients. For simplicity of notation, it is assumed that the variances of all the fading coefficients are one, but the result holds for other values as well.

Let $E$ denote an error event that the transmitted message pair $(x_A,x_B)$ is wrongly decoded at D. 

The probability of $E$ conditioned on $H$ given in \eqref{pe_H}, can be upper bounded as in \eqref{pe_H_ub} (eqns. \eqref{pe_H} and \eqref{pe_H_ub} are shown at the top of the next page). $P_{H}\left \lbrace f(\hat{x}_A^R, \hat{x}_B^R) = f\left(x_A,x_B\right)\right \rbrace$ and $P_{H}\left \lbrace f(\hat{x}_A^R, \hat{x}_B^R) \neq f\left(x_A,x_B\right)\right \rbrace$ respectively denote the probabilities that $R$ transmits the correct and wrong network coded symbol during Phase 2, for a given $H.$ Also, $P_{H}\left \lbrace(\hat{x}_A^D , \hat{x}_B^D) \neq \left(x_A,x_B\right) \bigm\vert f(\hat{x}_A^R, \hat{x}_B^R)= f\left(x_A,x_B\right)\right\rbrace$ and $P_{H}\left \lbrace(\hat{x}_A^D , \hat{x}_B^D) \neq \left(x_A,x_B\right) \bigm\vert f(\hat{x}_A^R, \hat{x}_B^R)\neq f\left(x_A,x_B\right)\right\rbrace$ respectively denote the probabilities of $E$ given that $R$ transmitted the correct and wrong network coded symbol for a given $H.$  $P_{H}\lbrace E \rbrace$ can be upper bounded as in \eqref{pe_H_ub_equal}, where $P_{H}\left \lbrace f(\hat{x}_A^R, \hat{x}_B^R) = x'_R\right \rbrace$ denotes the probability that the network coded symbol transmitted by $R$ is $x'_R \neq f(x_A,x_B)$ and $ P_{H}\left \lbrace(\hat{x}_A^D , \hat{x}_B^D) \neq \left(x_A,x_B\right) \bigm\vert f(\hat{x}_A^R, \hat{x}_B^R)= x'_R\right\rbrace$ is the probability of $E$ given that R transmits $x'_R,$ for a given $H.$ Taking expectation of the terms in \eqref{pe_H_ub_equal} w.r.t $H,$ we get \eqref{pe_H_ub_equal_avg}.

\begin{figure*}
\scriptsize
\begin{align}
\nonumber
P_{H}\lbrace E \rbrace &= P_{H}\left \lbrace(\hat{x}_A^D , \hat{x}_B^D) \neq \left(x_A,x_B\right) \bigm\vert f(\hat{x}_A^R, \hat{x}_B^R) = f\left(x_A,x_B\right)\right\rbrace P_{H}\left \lbrace f(\hat{x}_A^R, \hat{x}_B^R) = f\left(x_A,x_B\right)\right \rbrace\\
\label{pe_H}
&\hspace{5.5 cm}+ P_{H}\left \lbrace(\hat{x}_A^D , \hat{x}_B^D) \neq \left(x_A,x_B\right) \bigm\vert f(\hat{x}_A^R, \hat{x}_B^R) \neq f\left(x_A,x_B\right)\right\rbrace P_{H}\left \lbrace f(\hat{x}_A^R, \hat{x}_B^R) \neq f\left(x_A,x_B\right)\right \rbrace\\
\label{pe_H_ub}
&\leq P_{H}\left \lbrace(\hat{x}_A^D , \hat{x}_B^D) \neq \left(x_A,x_B\right) \bigm\vert f(\hat{x}_A^R, \hat{x}_B^R)= f\left(x_A,x_B\right)\right\rbrace + P_{H}\left \lbrace(\hat{x}_A^D , \hat{x}_B^D) \neq \left(x_A,x_B\right) \bigm\vert f(\hat{x}_A^R, \hat{x}_B^R) \neq f\left(x_A,x_B\right)\right\rbrace P_{H}\left \lbrace f(\hat{x}_A^R, \hat{x}_B^R) \neq f\left(x_A,x_B\right)\right \rbrace\\
\label{pe_H_ub_equal}
& \leq  P_{H}\left \lbrace(\hat{x}_A^D , \hat{x}_B^D) \neq \left(x_A,x_B\right) \bigm\vert f(\hat{x}_A^R, \hat{x}_B^R)= f\left(x_A,x_B\right)\right\rbrace + \sum_{\substack{{f(\hat{x}_A^R, \hat{x}_B^R) = x'_R,}\\{x'_R \neq f(x_A,x_B)}}}P_{H}\left \lbrace(\hat{x}_A^D , \hat{x}_B^D) \neq \left(x_A,x_B\right) \bigm\vert f(\hat{x}_A^R, \hat{x}_B^R)= x'_R\right\rbrace P_{H}\left \lbrace f(\hat{x}_A^R, \hat{x}_B^R) = x'_R\right \rbrace\\
\label{pe_H_ub_equal_avg}
P\lbrace E \rbrace &\leq P\left \lbrace(\hat{x}_A^D , \hat{x}_B^D) \neq \left(x_A,x_B\right) \bigm\vert f(\hat{x}_A^R, \hat{x}_B^R)= f\left(x_A,x_B\right)\right\rbrace + \hspace{-.5 cm}\sum_{\substack{{f(\hat{x}_A^R, \hat{x}_B^R) = x'_R,}\\{x'_R \neq f(x_A,x_B)}}}\hspace{-.5 cm}P\left \lbrace(\hat{x}_A^D , \hat{x}_B^D) \neq \left(x_A,x_B\right) \bigm\vert f(\hat{x}_A^R, \hat{x}_B^R)= x'_R\right\rbrace P\left \lbrace f(\hat{x}_A^R, \hat{x}_B^R) = x'_R\right \rbrace.
\end{align}
\hrule
\end{figure*}
The rest of the proof of Theorem 1 is presented in two parts as Lemma 1 and Lemma 2. In Lemma 1, it is shown that $P\left \lbrace(\hat{x}_A^D , \hat{x}_B^D) \neq \left(x_A,x_B\right) \bigm\vert f(\hat{x}_A^R, \hat{x}_B^R)= f\left(x_A,x_B\right)\right\rbrace $ has a diversity order two. Lemma 2 shows that $P\left \lbrace(\hat{x}_A^D , \hat{x}_B^D) \neq \left(x_A,x_B\right) \bigm\vert f(\hat{x}_A^R, \hat{x}_B^R)= x'_R\right\rbrace$ has a diversity order one. Since $P\left \lbrace f(\hat{x}_A^R, \hat{x}_B^R) = x'_R\right \rbrace$ has a diversity order one, Lemma 1 and Lemma 2 together imply that $P\left \lbrace E \right\rbrace$ has a diversity order two.

\begin{lemma}
 The probability $P\left \lbrace(\hat{x}_A^D , \hat{x}_B^D) \neq \left(x_A,x_B\right)\bigm\vert\right.$ $\left. f(\hat{x}_A^R, \hat{x}_B^R)= f\left(x_A,x_B\right)\right\rbrace $ has a diversity order two.
\begin{proof}

Recall that the decoder used at D given in \eqref{eqn_decoder} in Section II A, involves computation of the metrics $m_1$ and $m_2$ defined in \eqref{metric_m1} and \eqref{metric_m2}. Under the condition that $R$ transmitted the correct network coding symbol, a decoding error occurs at $D$ only when $m_1(x_A,x_B)>m_1(x'_A,x'_B)$ or $m_1(x_A,x_B)>\log(SNR)+m_2(x'_A,x'_B)$ for some $(x'_A,x_B) \neq (x_A,x_B).$ Hence $P\left \lbrace(\hat{x}_A^D, \hat{x}_B^D) \neq \left(x_A,x_B\right) \bigm\vert f(\hat{x}_A^R, \hat{x}_B^R)= f\left(x_A,x_B\right)\right\rbrace$ can be upper bounded as in \eqref{pe_latest_ub}, which can be upper bounded using the union bound as in \eqref{p1_H_ub2} (eqns. \eqref{pe_latest_ub} and \eqref{p1_H_ub2} are given at the top of the next page).

{\begin{figure*}
\scriptsize
\begin{align}
\nonumber
P\left \lbrace(\hat{x}_A^D, \hat{x}_B^D) \neq \left(x_A,x_B\right) \bigm\vert f(\hat{x}_A^R, \hat{x}_B^R)= f\left(x_A,x_B\right)\right\rbrace
&=P \left \lbrace \left \lbrace m_1(x_A,x_B) > m_1(x'_A,x'_B), (x'_A,x'_B) \neq (x_A,x_B))\right \rbrace \right.\\
\label{pe_latest_ub}
& \left.\hspace{-.8 cm}\cup \left \lbrace m_1(x_A,x_B) > \log(SNR)
+m_2(x'_A,x'_B), (x'_A,x'_B) \neq (x_A,x_B) \right \rbrace \bigm\vert f(\hat{x}_A^R, \hat{x}_B^R)= f\left(x_A,x_B\right)\right \rbrace\\
\nonumber
& \hspace{-0 cm}\leq\hspace{-.7 cm}\sum_{\substack{{(x'_A,x'_B) \in \mathcal{S}^2}\\{(x'_A,x'_B)\neq (x_A,x_B)}}} \hspace{-.7 cm}P \left \lbrace m_1(x_A,x_B)> m_1(x'_A,x'_B)\bigm\vert f(\hat{x}_A^R, \hat{x}_B^R)= f\left(x_A,x_B\right)\right \rbrace \\
\label{p1_H_ub2}
&\hspace{.7 cm}+ \hspace{-0.7 cm}\sum_{\substack{{(x'_A,x'_B) \in \mathcal{S}^2}\\{(x'_A,x'_B)\neq (x_A,x_B)}}}\hspace{-.7 cm} P\left \lbrace m_1(x_A,x_B)> \log(SNR)+m_2(x'_A,x'_B)\bigm\vert f(\hat{x}_A^R, \hat{x}_B^R)= f\left(x_A,x_B\right)\right \rbrace.
\end{align}
\hrule
\end{figure*}
}

 $P\left \lbrace m_1(x_A,x_B)> m_1(x'_A,x'_B) \bigm\vert f(\hat{x}_A^R, \hat{x}_B^R)= f\left(x_A,x_B\right)\right \rbrace$ is equal to the Pair-wise Error Probability (PEP) of a space time coded $3 \times 1$ collocated MISO system, with the codeword difference matrices of the space time code used at the transmitter being of the form {\footnotesize $\begin{bmatrix} a \Delta x_A & c \Delta x_A \\ b \Delta x_B & d \Delta x_B \\0 & f(x_A,x_B)-f(x'_A,x'_B)\end{bmatrix},$} where $\Delta x_A=x_A-x'_A, \Delta x_B= x_B-x'_B.$ When $\Delta x_A, \Delta x_B \neq 0,$ these codeword difference matrices are of rank 2, since the restricted codeword difference matrices are full rank. When $\Delta x_A =0$ and $\Delta x_B \neq 0$ (and also $\Delta x_B =0,\Delta x_A \neq 0$), the codeword difference matrices are full rank, since $f(x_A,x_B)\neq f(x_A,x'_B)$ (otherwise the exclusive law given in \eqref{ex_law} will be violated). Since the codeword difference matrices are full rank, the probability {\footnotesize$P\left \lbrace m_1(x_A,x_B)> m_1(x'_A,x'_B) \bigm\vert f(\hat{x}_A^R, \hat{x}_B^R)= f\left(x_A,x_B\right)\right \rbrace$} has a diversity order two \cite{TaSeCa}.
\begin{figure*}
\scriptsize
\begin{align}
\label{metric_m4}
&m_4(x_A,x_B,x_R)=\left\vert y_{D_1}-h_{AD} \sqrt{E_s} a \: x_A- h_{BD} \sqrt{E_s} b \: x_B\right\vert^2+ \left\vert y_{D_2} -h_{AD} \sqrt{E_s} c \: x_A- h_{BD} \sqrt{E_s} d \: x_B- h_{RD} \sqrt{E_s} x_R \right\vert ^2+\log(SNR).
\end{align}
\hrule
\end{figure*}
{
\begin{figure*}
\scriptsize
\begin{align}
\label{eqn_m1_m4}
P \left \lbrace m_1(x_A,x_B)> m_2(x'_A,x'_B)+\log(SNR) \bigm\vert f(\hat{x}_A^R, \hat{x}_B^R)= f\left(x_A,x_B\right)\right \rbrace &=P \left \lbrace m_1(x_A,x_B)> \hspace{-.5 cm}\min_{x'_R \neq f(x'_A,x'_B)} m_4(x'_A,x'_B,x'_R) \bigm\vert f(\hat{x}_A^R, \hat{x}_B^R)= f\left(x_A,x_B\right)\right \rbrace\\
\label{p1_ub4}
&\leq\hspace{-.5 cm} \sum_{x'_R \neq f(x'_A,x'_B)} \hspace{-.5 cm}P \left \lbrace m_1(x_A,x_B)> m_4(x'_A,x'_B,x'_R) \bigm\vert f(\hat{x}_A^R, \hat{x}_B^R)= f\left(x_A,x_B\right)\right \rbrace\\
\hline
\nonumber
&\hspace{-9 cm}P_H \left \lbrace m_1(x_A,x_B)>m_4(x'_A,x'_B,x'_R)\bigm\vert f(\hat{x}_A^R, \hat{x}_B^R) = f(x_A,x_B)\right \rbrace\\
\label{pe_eqn_first}
&\hspace{-9cm}= P_H\left \lbrace \right \vert z_{D_1}\vert^2+\vert z_{D_2}\vert^2
 > \log(SNR)+\vert z_{D_1}+h_{AD}\sqrt{E_s}a\Delta x_A +h_{BD}\sqrt{E_s}b\Delta x_B\vert^2+\vert z_{D_2}+ h_{AD}\sqrt{E_s}c \Delta x_A+h_{BD}\sqrt{E_s}d \Delta x_B +h_{RD}\sqrt{E_s}\Delta x_R\vert^2\rbrace.\\
\label{pe_eqn_1_first}
&\hspace{-9cm}=P_{H}\left\lbrace 2Re\left\lbrace \mathbf{z_D}^*\frac{\mathbf{\texttt{x}}}{\left \|\mathbf{\texttt{x}}\right \|}\right \rbrace \leq \frac{-\log\left(SNR\right)}{\left\|\mathbf{\texttt{x}}\right\|}-{\left\|\mathbf{\texttt{x}}\right\|}\right\rbrace=P_{H}\left\lbrace w \leq \frac{-\log\left(SNR\right)}{\sqrt{2}\left\|\mathbf{\texttt{x}}\right\|}-\frac{\left\|\mathbf{\texttt{x}}\right\|}{\sqrt{2}}\right\rbrace.
\end{align}
\hrule
\end{figure*}
} 
Let $m_4$ be a metric as defined in \eqref{metric_m4}, shown at the top of this page. The probability $P \left \lbrace m_1(x_A,x_B)> m_2(x'_A,x'_B)+\log(SNR)\right. \left.\bigm\vert\right.$ $\left.f(\hat{x}_A^R, \hat{x}_B^R)= f\left(x_A,x_B\right)\right \rbrace$ can be written in terms of the metrics $m_1$ and $m_4$ as in \eqref{eqn_m1_m4}, which can be upper bounded as in \eqref{p1_ub4} (eqns. \eqref{eqn_m1_m4} --  \eqref{pe_eqn_1_first} are shown at the top of this page).

Let $\Delta x_R=f(x_A,x_B)-x'_R,$ $\Delta x_A=x_A-x'_A$  and $\Delta x_B=x_B-x'_B.$ 
The probability {\small $P_H \left \lbrace m_1(x_A,x_B)>m_4(x'_A,x'_B,x'_R)\bigm\vert f(\hat{x}_A^R, \hat{x}_B^R) = f(x_A,x_B)\right \rbrace$} can be written in terms of the additive noise $z_{D_1}$ and $z_{D_2},$ as given in \eqref{pe_eqn_first}.

Let $x_1=(h_{AD}a\Delta x_A + h_{BD} b \Delta x_B)\sqrt{E_s},$ $x_2=(h_{AD}c\Delta x_A + h_{BD} d \Delta x_B+h_{RD} \Delta x_R)\sqrt{E_s}.$ Also, let $\mathbf{z_D}=[z_{D_1}\; z_{D_2}]$ and $\mathbf{\texttt{x}}=[x_1 \; x_2]^T.$
Then \eqref{pe_eqn_first} can be simplified as in \eqref{pe_eqn_1_first}, where $w=\sqrt{2}Re\{\mathbf{z_D^*}\frac{\mathbf{\texttt{x}}}{\|\mathbf{\texttt{x}} \|}\},$ is distributed according to $\mathcal{N}(0,1).$  In terms of the $Q$ function, the probability $P_{H}\left\lbrace w \leq \frac{-\log\left(SNR\right)}{\sqrt{2}\left\|\mathbf{\texttt{x}}\right\|}-\frac{\left\|\mathbf{\texttt{x}}\right\|}{\sqrt{2}}\right\rbrace$ in \eqref{pe_eqn_1_first} can be written as $Q \left[ \frac{\log\left(SNR\right)}{\sqrt{2}\left\|\mathbf{\texttt{x}}\right\|}+\frac{\left\|\mathbf{\texttt{x}}\right\|}{\sqrt{2}}\right].$ Note that $\left\|\mathbf{\texttt{x}}\right\|$ depends on the fade coefficients. To complete the proof, it suffices to show that $\mathbb{E}\left(Q \left[ \frac{\log\left(SNR\right)}{\sqrt{2}\left\|\mathbf{\texttt{x}}\right\|}+\frac{\left\|\mathbf{\texttt{x}}\right\|}{\sqrt{2}}\right]\right)$ has a diversity order two.

The vector $\mathbf{\texttt{x}}$ can be written as,
$
\mathbf{\texttt{x}}=\sqrt{E_s}\underbrace{[h_{AD}\; h_{BD}\; h_{RD}]}_{\mathbf{h}}\underbrace{\begin{bmatrix}a \Delta x_A & c \Delta x_A\\b\Delta x_B & d\Delta x_B\\0 &\Delta x_R\end{bmatrix}}_{\mathbf{\Delta X}}.
$

Since $\mathbf{\Delta X \Delta X^*}$ is Hermitian, it is unitarily diagonalizable, i.e, $ \mathbf{\Delta X \Delta X^*}=\mathbf{U}\mathbf{\Sigma}\mathbf{U^*},$ where $\mathbf{U}$ is unitary and $\mathbf{\Sigma}=\begin{bmatrix} \lambda_1 & 0 & 0\\0 & \lambda_2 & 0\\0 & 0 & 0 \end{bmatrix}$ with $\lambda_1 \geq \lambda_2.$  We have $\left\|\mathbf{\texttt{x}}\right\|^2=SNR\;\mathbf{h}\mathbf{U}\mathbf{\Sigma}\mathbf{U}^*\mathbf{h}^*.$ Let $\mathbf{\tilde{h}}=\mathbf{h}\mathbf{U}=[\tilde{h}_1 \; \tilde{h}_2 \; \tilde{h}_3].$ The vector $\mathbf{\tilde{h}}$ has the same distribution as that of $\mathbf{h},$ since $\mathbf{U}$ is unitary.

Since the rank of $\mathbf{\Delta X}$ is at least one, $\lambda_1>0.$ We consider the two cases where $\lambda_2>0$ and $\lambda_2=0.$\\
\textit{\textbf{\noindent \hspace{-.5  cm}Case 1:}} $\lambda_2>0$\\
For this case, upper bounding $Q \left[ \frac{\log\left(SNR\right)}{\sqrt{2}\left\|\mathbf{\texttt{x}}\right\|}+\frac{\left\|\mathbf{\texttt{x}}\right\|}{\sqrt{2}}\right]$ by $Q \left[\frac{\left\|\mathbf{\texttt{x}}\right\|}{\sqrt{2}}\right],$ which is upper bounded by $e^{-\frac{\left\|\mathbf{\texttt{x}}\right\|^2}{4}},$ we have
\begin{align}
\label{eqn_eqn1}
\hspace{-.29 cm}Q \left[ \frac{\log\left(SNR\right)}{\sqrt{2}\left\|\mathbf{\texttt{x}}\right\|}+\frac{\left\|\mathbf{\texttt{x}}\right\|}{\sqrt{2}}\right] \leq e^{-\frac{1}{4}(\lambda_1 SNR \vert \tilde{h}_1 \vert^2+\lambda_2 SNR  \vert \tilde{h}_2 \vert^2)}.
\end{align}
Taking expectation w.r.t $\vert \tilde{h}_1 \vert$ and $\vert \tilde{h}_2 \vert,$ from \eqref{eqn_eqn1}, we get, 
  $\mathbb{E}\left(Q \left[ \frac{\log\left(SNR\right)}{\sqrt{2}\left\|\mathbf{\texttt{x}}\right\|}+\frac{\left\|\mathbf{\texttt{x}}\right\|}{\sqrt{2}}\right]\right) \leq \frac{1}{\left(1+\frac{\lambda_1 SNR}{4}\right)\left(1+\frac{\lambda_2 SNR}{4}\right)}.$ Hence $\mathbb{E}\left(Q \left[ \frac{\log\left(SNR\right)}{\sqrt{2}\left\|\mathbf{\texttt{x}}\right\|}+\frac{\left\|\mathbf{\texttt{x}}\right\|}{\sqrt{2}}\right]\right)$ has a diversity order two.
  
\textit{\textbf{\noindent \hspace{-.5  cm}Case 2:}} $\lambda_2=0$\\
 For this case $ \left\|\mathbf{\texttt{x}}\right\|=\sqrt{\lambda_1 SNR} \vert \tilde{h}_1\vert.$ Hence, 
 \begin{align}
 \nonumber
 Q \left[ \frac{\log\left(SNR\right)}{\sqrt{2}\left\|\mathbf{\texttt{x}}\right\|}+\frac{\left\|\mathbf{\texttt{x}}\right\|}{\sqrt{2}}\right]&\\
 \label{eqn_eqn2} 
 &\hspace{-2.5 cm}=Q \left[ \frac{\log\left(SNR\right)}{\sqrt{2}\sqrt{\lambda_1 SNR }\vert \tilde{h}_1\vert}+\frac{\sqrt{\lambda_1 SNR }\vert \tilde{h}_1\vert}{\sqrt{2}}\right].
 \end{align}

Let $r=\vert \tilde{h}_1\vert.$ Taking expectation w.r.t $r,$ from \eqref{eqn_eqn2}, we get,

{\small
\begin{align}
\nonumber
\mathbb{E}\left(Q \left[ \frac{\log\left(SNR\right)}{\sqrt{2}\left\|\mathbf{\texttt{x}}\right\|}+\frac{\left\|\mathbf{\texttt{x}}\right\|}{\sqrt{2}}\right]\right)&\\
\nonumber
&\hspace{-3.8 cm}=\int_{r=0}^{\infty} Q \left[ \frac{\log\left(SNR\right)}{\sqrt{2}\sqrt{\lambda_1 SNR }r}+\frac{\sqrt{\lambda_1 SNR }r}{\sqrt{2}}\right] 2r e^{-{r^2}} dr\\
\nonumber
&\hspace{-3.8 cm}=\underbrace{\int_{r=0}^{\sqrt{\frac{\log(SNR)}{\lambda_1 SNR}}} 2Q \left[ \frac{\log\left(SNR\right)}{\sqrt{2}\sqrt{\lambda_1 SNR }r}+\frac{\sqrt{\lambda_1 SNR }r}{\sqrt{2}}\right] r e^{-{r^2}} dr}_{I_1}\\
\nonumber
 &\hspace{-3.5 cm}+\underbrace{\int_{r=\sqrt{\frac{\log(SNR)}{\lambda_1 SNR}}}^{\infty} 2Q \left[ \frac{\log\left(SNR\right)}{\sqrt{2}\sqrt{\lambda_1 SNR }r}+\frac{\sqrt{\lambda_1 SNR }r}{\sqrt{2}}\right] r e^{-{r^2}} dr}_{I_2}.
\end{align}}

In the following, we show that the integrals $I_1$ and $I_2$ have diversity order two. Note that $ \frac{\log\left(SNR\right)}{\sqrt{2}\sqrt{\lambda_1 SNR }r}+\frac{\sqrt{\lambda_1 SNR }r}{\sqrt{2}},$ as a function of $r,$ attains the minimum value when $r={\sqrt{\frac{\log(SNR)}{\lambda_1 SNR}}}$ and the minimum value equals $\sqrt{2\log(SNR)}.$ Since, $Q(x)$ is a decreasing function of $x,$ we have, $Q\left[\frac{\log\left(SNR\right)}{\sqrt{2}\sqrt{\lambda_1 SNR }r}+\frac{\sqrt{\lambda_1 SNR }r}{\sqrt{2}}\right]\leq Q \left[\sqrt{2\log(SNR)}\right].$ Hence, we have,
\begin{align*}
I_1 &\leq 2Q\left[\sqrt{2\log(SNR)}\right] \int_{r=0}^{\sqrt{\frac{\log(SNR)}{\lambda_1 SNR}}} r e^{-{r^2}} dr\\
&\leq \frac{2}{SNR}\left(1-e^{-\frac{\log(SNR)}{\lambda_1 SNR}}\right).
\end{align*}

Since for small $x,$ $e^{-x}$ can be approximated as $1-x,$ at high $SNR,$ we have 
$
I_1 \leq\frac{2}{SNR}\frac{\log(SNR)}{\lambda_1 SNR}
.$ Since $\displaystyle{\lim_{SNR \rightarrow \infty} \frac{-\log \left(\frac{2\log(SNR)}{\lambda_1 {SNR}^2}\right)}{\log(SNR)}=2},$ $I_1$ has a diversity order at least two.

Let $r_0=\sqrt{\frac{\log(SNR)}{\lambda_1 SNR}}.$
The integral $I_2$ can be upper bounded as,
$I_2\leq{\int_{r=r_0}^{\infty} Q \left[ \frac{\log\left(SNR\right)}{\sqrt{2}\sqrt{\lambda_1 SNR }r}+\frac{\sqrt{\lambda_1 SNR }r}{\sqrt{2}}\right] r  dr}.
$ Let $r'=\frac{\log\left(SNR\right)}{\sqrt{2}\sqrt{\lambda_1 SNR }r}+\frac{\sqrt{\lambda_1 SNR }r}{\sqrt{2}}.$ 
As a function of $r,$ $r'$ is monotonically increasing for $r \geq r_0.$ Also, for $r \geq r_0,$ $r$ can be written in terms of $r'$ as, $r=\frac{r'+\sqrt{{r'}^2-2\log(SNR)}}{\sqrt{2 \lambda_1 SNR}}.$ 
We have, $d {r}= d{r'} \frac{1}{\sqrt{2\lambda_1 SNR}} \left(1+\frac{r'}{\sqrt{{r'}^2-2\log(SNR)}}\right).$ Since $r \leq \frac{2r'}{\sqrt{2 \lambda_1 SNR}},$
$I_2$ can be upper bounded in terms of $r'$ as,

{\small
\begin{align*}
I_2&\leq{\int_{\sqrt{2\log(SNR)}}^{\infty}}  \frac{Q(r')r'}{{ \lambda_1 SNR}}  \left(1+\frac{r'}{\sqrt{{r'}^2-2\log(SNR)}}\right) d{r'}\\
&=\underbrace{\frac{1}{{\lambda_1 SNR}}\int_{\sqrt{2\log(SNR)}}^{\infty} Q(r') r' d{r'}}_{I_{21}}\\
&\hspace{1.6 cm}+\underbrace{\frac{1}{ \lambda_1 SNR}\int_{\sqrt{2\log(SNR)}}^{\infty} \frac{Q(r'){r'}^2}{\sqrt{{r'}^2-2\log(SNR)}} d {r'}}_{I_{22}}.
\end{align*}}

Upper bounding $Q(r')$ by $e^{-\frac{r'^2}{2}},$ $I_{21}$ can be shown to be upper bounded as $\frac{2}{\lambda_1 SNR^2},$ which falls as ${SNR}^{-2}.$ Upper bounding $Q(r')$ by $e^{-\frac{r'^2}{2}},$ and using the transformation $t={r'}^2-2\log(SNR),$ $I_{22}$ can be upper bounded as,
\begin{align*}
I_{22} &\leq \frac{1}{\lambda_1 SNR^2}\int_{0}^{\infty} \frac{t e^{-\frac{t}{2}}}{\sqrt{t}{\sqrt{t+2\log(SNR)}}} dt \\
&  \hspace{1.8 cm} + \frac{2\log(SNR)}{\lambda_1 SNR^2}\int_{0}^{\infty} \frac{ e^{-\frac{t}{2}}}{\sqrt{t}{\sqrt{t+2\log(SNR)}}} dt\\
& \leq \frac{1}{\lambda_1 SNR^2}\int_{0}^{\infty} e^{-\frac{t}{2}} dt + \frac{1}{\lambda_1 SNR^2}\int_{0}^{\infty} \frac{e^{-\frac{t}{2}}}{\sqrt{t}} dt\\
&=\frac{2+2\sqrt{2\pi}\log(SNR)}{\lambda_1 SNR^2}.
\end{align*}
where the second inequality above follows from the facts that $\frac{1}{\sqrt{t+2 \log(SNR)}} \leq \frac{1}{\sqrt{t}}$ and $\frac{1}{\sqrt{t+2 \log(SNR)}} \leq 1$ for sufficiently large $SNR.$ The last equality follows from the fact that $\int_{0}^{\infty} \frac{e^{-\frac{t}{2}}}{\sqrt{t}} dt=\sqrt{2}\Gamma(1/2)=\sqrt{2\pi},$ where $\Gamma(z)$ is the integral, $\Gamma(z)=\int_{0}^{\infty} e^{-t} t^{z-1} dt.$ Since, $\displaystyle\lim_{SNR \rightarrow \infty} \frac{-\log \left(\frac{2+2\sqrt{2\pi}\log(SNR)}{\lambda_1 SNR^2}\right)}{\log(SNR)}=2,$ $I_{22}$ has a diversity order 2. This completes the proof of Lemma 1.
\end{proof}
\end{lemma}

\begin{lemma}
The probability $P\left \lbrace(\hat{x}_A^D , \hat{x}_B^D) \neq \left(x_A,x_B\right) \bigm\vert \right.$ $\left.f(\hat{x}_A^R, \hat{x}_B^R)= x'_R\right\rbrace, x'_R \neq f(x_A,x_B),$ has a diversity order one.
\begin{proof}
\begin{figure*}
\scriptsize
\begin{align}
\label{metric_m4_repeat}
&m_4(x_A,x_B,x_R)=\left\vert y_{D_1}-h_{AD} \sqrt{E_s} a \: x_A- h_{BD} \sqrt{E_s} b \: x_B\right\vert^2+ \left\vert y_{D_2} -h_{AD} \sqrt{E_s} c \: x_A- h_{BD} \sqrt{E_s} d \: x_B- h_{RD} \sqrt{E_s} x_R \right\vert ^2+\log(SNR).
\end{align}
\hrule
\end{figure*}
Let $m_4$ denote the metric as defined in \eqref{metric_m4_repeat}, shown at the top of the next page.
Under the condition that $R$ transmitted the wrong network coded symbol $x'_R,$ a decoding error occurs at $D$ only when $m_4(x_A,x_B,x'_R)>m_4(x''_A,x''_B,x''_R)$ or $m_4(x_A,x_B,x'_R)>m_1(x''_A,x''_B),$ for some $(x''_A,x''_B) \neq (x_A,x_B)$ and $x''_R \neq f(x''_A,x''_B).$ Hence, $P\left \lbrace(\hat{x}_A^D , \hat{x}_B^D) \neq \left(x_A,x_B\right) \bigm\vert f(\hat{x}_A^R, \hat{x}_B^R) = x'_R\right\rbrace$ can be upper bounded as in \eqref{pe_latest_ub_lemma2} (eqns. \eqref{pe_latest_ub_lemma2} -- \eqref{pe_eqn_1} are shown at the top of the next page). Using the union bound, from \eqref{pe_latest_ub_lemma2}, we get \eqref{pe_latest_ub1_lemma2}.

{\begin{figure*}
\scriptsize
\begin{align}
\nonumber
P\left \lbrace(\hat{x}_A^D , \hat{x}_B^D) \neq \left(x_A,x_B\right) \bigm\vert f(\hat{x}_A^R, \hat{x}_B^R) = x'_R\right\rbrace
&=\hspace{-.7 cm}\sum_{\substack{{(x''_A,x''_B) \in \mathcal{S}^2}\\{(x''_A,x''_B)\neq (x_A,x_B)}}} \hspace{-.7 cm} P \left \lbrace \left \lbrace m_4(x_A,x_B,x'_R) > m_4(x''_A,x''_B,x''_R), x''_R \neq f(x''_A,x''_B)\right \rbrace \right.\\
\label{pe_latest_ub_lemma2}
&\left.
\hspace{4.5 cm}\cup \left \lbrace m_4(x_A,x_B,x'_R) > 
m_1(x''_A,x''_B) \right \rbrace \bigm\vert f(\hat{x}_A^R, \hat{x}_B^R)= x'_R\right \rbrace\\
\nonumber
&\leq \hspace{-.4 cm}\sum_{\substack{{(x''_A,x''_B) \in \mathcal{S}^2}\\{(x''_A,x''_B)\neq (x_A,x_B)}}} \hspace{-.0cm}\sum_{\substack{{x''_R \in \mathcal{S},}\\{ x''_R \neq f(x''_A,x''_B)}}}\hspace{-.3 cm} P \left \lbrace m_4(x_A,x_B,x'_R)>m_4(x''_A,x''_B,x''_R)\right.\left.\bigm\vert f(\hat{x}_A^R, \hat{x}_B^R) = x'_R\right\rbrace\\
\label{pe_latest_ub1_lemma2}
&\hspace{2 cm}+\hspace{-.57 cm}\sum_{\substack{{(x''_A,x''_B) \in \mathcal{S}^2}\\{(x''_A,x''_B)\neq (x_A,x_B)}}}\hspace{-.5 cm} P \hspace{-.1 cm}\left \lbrace m_4(x_A,x_B,x'_R)>m_1(x''_A,x''_B)\right. \left.\bigm\vert f(\hat{x}_A^R, \hat{x}_B^R) = x'_R\right\rbrace.
\end{align}
\hrule
\end{figure*}}

Since the matrix $\begin{bmatrix} a (x_A-x''_A) & c (x_A-x''_A) \\ b (x_B-x''_B) & d (x_B-x''_B) \\0 & x'_R-x''_R\end{bmatrix},$
has rank at least one for $(x_A,x_B) \neq (x''_A,x''_B),$ $P \left \lbrace m_4(x_A,x_B,x'_R)>m_4(x''_A,x''_B,x''_R)\bigm\vert f(\hat{x}_A^R, \hat{x}_B^R) = x'_R\right \rbrace$ has a diversity order at least one.
 
Let $\Delta x_R=x'_R-f(x''_A,x''_B),$ $\Delta x_A=x_A-x'_A$  and $\Delta x_B=x_B-x'_B.$ 
The probability {\small$P_H \left \lbrace m_4(x_A,x_B,x'_R)>m_1(x''_A,x''_B)\bigm\vert f(\hat{x}_A^R, \hat{x}_B^R) = x'_R\right \rbrace$} can be written in terms of the additive noise $z_{D_1}$ and $z_{D_2},$ as given in \eqref{pe_eqn}.

{\begin{figure*}
\scriptsize
\begin{align}
\nonumber
P_H \left \lbrace \log(SNR)+m_4(x_A,x_B,x'_R)>m_1(x''_A,x''_B)\bigm\vert f(\hat{x}_A^R, \hat{x}_B^R) = x'_R\right \rbrace\\
\label{pe_eqn}
&\hspace{-8cm}= P_H\left \lbrace \right \vert z_{D_1}\vert^2+\vert z_{D_2}\vert^2+\log(SNR)
 > \vert z_{D_1}+h_{AD}\sqrt{E_s}a\Delta x_A +h_{BD}\sqrt{E_s}b\Delta x_B\vert^2+\vert z_{D_2}+h_{AD}\sqrt{E_s}c\Delta x_A+h_{BD}\sqrt{E_s}d \Delta x_B +h_{RD}\Delta x_R\vert^2\rbrace.\\
\label{pe_eqn_1}
&\hspace{-8cm}=P_{H}\left\lbrace 2Re\left\lbrace \mathbf{z_D}^*\frac{\mathbf{\texttt{x}}}{\left \|\mathbf{\texttt{x}}\right \|}\right \rbrace \leq \frac{\log\left(SNR\right)}{\left\|\mathbf{\texttt{x}}\right\|}-{\left\|\mathbf{\texttt{x}}\right\|}\right\rbrace=P_{H}\left\lbrace w \leq \frac{\log\left(SNR\right)}{\sqrt{2}\left\|\mathbf{\texttt{x}}\right\|}-\frac{\left\|\mathbf{\texttt{x}}\right\|}{\sqrt{2}}\right\rbrace.
\end{align}
\hrule
\end{figure*}}

Let $x_1=(h_{AD}a\Delta x_A + h_{BD} b \Delta x_B)\sqrt{E_s},$ $x_2=(h_{AD}c\Delta x_A + h_{BD} d \Delta x_B+h_{RD} \Delta x_R)\sqrt{E_s}.$ Also, let $\mathbf{z_D}=[z_{D_1}\; z_{D_2}]$ and $\mathbf{\texttt{x}}=[x_1 \; x_2]^T.$ Then \eqref{pe_eqn} can be simplified as in \eqref{pe_eqn_1},
 where $w=\sqrt{2}Re\{\mathbf{z_D^*}\frac{\mathbf{\texttt{x}}}{\|\mathbf{\texttt{x}} \|}\},$ is distributed according to $\mathcal{N}(0,1).$ Hence, 
{
\begin{align}
\nonumber
&P_{H}\hspace{-.1 cm}\left \lbrace  w \leq \frac{\log\left(SNR\right)}{\sqrt{2}\left\|\mathbf{\texttt{x}}\right\|}-\frac{{\left\|\mathbf{\texttt{x}}\right\|}}{\sqrt{2}}\right\rbrace\\
\nonumber
&= 1_{\{{\left\|\mathbf{\texttt{x}}\right\|^2}\leq \log\left(SNR\right)\}}\left(1-Q\left[\frac{\log\left(SNR\right)}{\sqrt{2}\left\|\mathbf{\texttt{x}}\right\|}-\frac{\left\|\mathbf{\texttt{x}}\right\|}{\sqrt{2}}\right]\right)\\
\label{pe_eqn2}
&\hspace{1.05 cm}+1_{\{{\left\|\mathbf{\texttt{x}}\right\|^2}> \log\left(SNR\right)\}}Q\left[-\frac{\log\left(SNR\right)}{\sqrt{2}\left\|\mathbf{\texttt{x}}\right\|}+\frac{\left\|\mathbf{\texttt{x}}\right\|}{\sqrt{2}}\right].
\end{align}}
Taking expectation with respect to the fade coefficients in \eqref{pe_eqn2}, $P\left \lbrace w \leq \frac{\log\left(SNR\right)}{\sqrt{2}\left\|\mathbf{\texttt{x}}\right\|}-\frac{\left\|\mathbf{\texttt{x}}\right\|}{\sqrt{2}}\right\rbrace$ can be upper bounded as,
{
\begin{align}
\nonumber
&\hspace{-0 cm}P\left \lbrace w \leq \frac{\log\left(SNR\right)}{\sqrt{2}\left\|\mathbf{\texttt{x}}\right\|}-\frac{\left\|\mathbf{\texttt{x}}\right\|}{\sqrt{2}}\right\rbrace \leq P\left \lbrace {\left\|\mathbf{\texttt{x}}\right\|}^2 \leq \log(SNR) \right \rbrace\\
\label{pe_eqn3}
&\hspace{0.2 cm}+\int_{H:\{{\left\|\mathbf{\texttt{x}}\right\|^2}> \log\left(SNR\right)\}} Q\left[-\frac{\log\left(SNR\right)}{\sqrt{2}\left\|\mathbf{\texttt{x}}\right\|}+\frac{\left\|\mathbf{\texttt{x}}\right\|}{\sqrt{2}}\right] dH.
\end{align}}

The vector $\mathbf{\texttt{x}}$ can be written as,
$
\mathbf{\texttt{x}}=\sqrt{E_s}\underbrace{[h_{AD}\; h_{BD}\; h_{RD}]}_{\mathbf{h}}\underbrace{\begin{bmatrix}a \Delta x_A & c \Delta x_A\\b\Delta x_B & d\Delta x_B\\0 &\Delta x_R\end{bmatrix}}_{\mathbf{\Delta X}}.
$ 

Since $\mathbf{\Delta X \Delta X^*}$ is Hermitian, it is unitarily diagonalizable, i.e, $ \mathbf{\Delta X \Delta X^*}=\mathbf{U}\mathbf{\Sigma}\mathbf{U^*},$ where $\mathbf{U}$ is unitary and $\mathbf{\Sigma}=\begin{bmatrix} \lambda_1 & 0 & 0\\0 & \lambda_2 & 0\\0 & 0 & 0 \end{bmatrix}$ with $\lambda_1 \geq \lambda_2.$ Since the rank of $\mathbf{\Delta X}$ is at least one, $\lambda_1>0.$ We have $\left\|\mathbf{\texttt{x}}\right\|^2=SNR\;\mathbf{h}\mathbf{U}\mathbf{\Sigma}\mathbf{U}^*\mathbf{h}^*.$ Let $\mathbf{\tilde{h}}=\mathbf{h}\mathbf{U}=[\tilde{h}_1 \; \tilde{h}_2 \; \tilde{h}_3].$ The vector $\mathbf{\tilde{h}}$ has the same distribution as that of $\mathbf{h},$ since $\mathbf{U}$ is unitary. Hence, we have,

{\small
\begin{align}
\nonumber 
P\left \lbrace {\left\|\mathbf{\texttt{x}}\right\|}^2 \leq \log(SNR) \right \rbrace&=P\left \lbrace \lambda_1 \vert \tilde{h}_1\vert^2+\lambda_2 \vert \tilde{h}_2\vert^2 \leq \frac{\log(SNR)}{SNR}\right \rbrace\\
\nonumber
& \leq P\left \lbrace \lambda_1 \vert \tilde{h}_1\vert^2 \leq \frac{\log(SNR)}{SNR}\right \rbrace.
\end{align}
}

Since, $\vert \tilde{h}_1 \vert ^2$ is exponentially distributed, 
{
\begin{align}
\nonumber 
P\left \lbrace {\left\|\mathbf{\texttt{x}}\right\|}^2 \leq \log(SNR) \right \rbrace \leq \left(1-e^{-\frac{\log(SNR)}{\lambda_1 SNR}}\right).
\end{align}}

At high SNR, $1-e^{-\frac{\log(SNR)}{\lambda_1 SNR}}$ can be approximated as $-\frac{\log(SNR)}{\lambda_1 SNR}.$  $P\left \lbrace {\left\|\mathbf{\texttt{x}}\right\|}^2 \leq \log(SNR) \right \rbrace$ has a diversity order at least one since  $\displaystyle{\lim_{SNR \rightarrow \infty} \frac{-\log\left(\frac{\log(SNR)}{\lambda_1 SNR}\right)}{\log \left(SNR \right)}=1.}$ 
Since $Q(x)<e^{-\frac{x^2}{2}},$ the integral on the right hand side of \eqref{pe_eqn3} can be upper bounded as,
{
\begin{align}
\nonumber
&\int_{H:\{{\left\|\mathbf{\texttt{x}}\right\|^2}> \log\left(SNR\right)\}} Q\left[-\frac{\log\left(SNR\right)}{\sqrt{2}\left\|\mathbf{\texttt{x}}\right\|}+\frac{\left\|\mathbf{\texttt{x}}\right\|}{\sqrt{2}}\right] dH \\
\label{eqn_temp}
&\leq \int_{H:\{{\left\|\mathbf{\texttt{x}}\right\|^2}> \log\left(SNR\right)\}} e^{-\frac{\left(-\frac{\log\left(SNR\right)}{\left\|\mathbf{\texttt{x}}\right\|}+{\left\|\mathbf{\texttt{x}}\right\|}\right)^2}{4}} dH.
\end{align}
} 
We consider the following two cases when $\lambda_2>0$ and $\lambda_2=0.$\\
\textit{\textbf{\hspace{-.7 cm}Case 1:}} $\lambda_2>0$\\
For this case, from the integral in \eqref{eqn_temp}, we get,
{
\begin{align}
\nonumber
&\int_{H:\{{\left\|\mathbf{\texttt{x}}\right\|^2}> \log\left(SNR\right)\}} Q\left[-\frac{\log\left(SNR\right)}{\sqrt{2}\left\|\mathbf{\texttt{x}}\right\|}+\frac{\left\|\mathbf{\texttt{x}}\right\|}{\sqrt{2}}\right] dH \\
\nonumber
&\leq \int_{\vert\tilde{h}_1\vert^2=0}^{\infty} \int_{\vert\tilde{h}_2\vert^2=0}^{\infty} e^{\log\left(SNR\right)} e^{-\frac{SNR(\lambda_1 \vert\tilde{h}_1\vert^2+\lambda_2 \vert\tilde{h}_2\vert^2)}{4}} \\
\nonumber
&\hspace{4.5 cm}e^{-\vert\tilde{h}_1\vert^2} e^{-\vert\tilde{h}_2\vert^2} d\vert\tilde{h}_1\vert^2\;d\vert\tilde{h}_2\vert^2\\
\nonumber
&=\frac{SNR}{\left({1+\frac{\lambda_1 SNR}{4}}\right)\left({1+\frac{\lambda_2 SNR}{4}}\right)}, 
\end{align}}which falls as $SNR^{-1}$ at high SNR.\\
\textit{\textbf{Case 2:}} $\lambda_2=0$
For this case,
{
\begin{align}
\nonumber
I \triangleq &\int_{H:\{{\left\|\mathbf{\texttt{x}}\right\|^2}> \log\left(SNR\right)\}} Q\left[-\frac{\log\left(SNR\right)}{\sqrt{2}\left\|\mathbf{\texttt{x}}\right\|}+\frac{\left\|\mathbf{\texttt{x}}\right\|}{\sqrt{2}}\right] f(H)dH \\
\nonumber
&\leq \int_{\vert\tilde{h}_1\vert >\frac{\sqrt{\log(SNR)}}{\sqrt{\lambda_1 \; SNR}}} Q\left[\frac{\sqrt{\lambda_1 SNR} \vert\tilde{h}_1\vert-\sqrt{\log(SNR)}}{\sqrt{2}}\right] \\
\label{eqn_temp2}
&\hspace{4.5 cm}2\vert \tilde{h}_1 \vert e^{-{\vert \tilde{h}_1 \vert^2}} d \vert\tilde{h}_1\vert.
\end{align}}
The above inequality follows from the fact that for $\vert\tilde{h}_1\vert >\frac{\sqrt{\log(SNR)}}{\sqrt{\lambda_1 \; SNR}},$ $-\frac{\log(SNR)}{\sqrt{\lambda_1 SNR}\vert \tilde{h}_1\vert}<-\sqrt{\log(SNR)}.$

Let $r=\vert \tilde{h}_1 \vert.$ From \eqref{eqn_temp2}, since $Q(x)<e^{-\frac{x^2}{2}}$ the integral $I$ can be upper bounded as,

{\vspace{-.7 cm}
\footnotesize
\begin{align}
\nonumber
&I \leq \int_{r >\underbrace{\frac{\sqrt{\log(SNR)}}{\sqrt{\lambda_1 \; SNR}}}_{r_0}} e^{-\underbrace{\left({1}+\frac{\lambda_1 SNR}{4}\right)}_{k_1}\left(r-\underbrace{\frac{\sqrt{\lambda_1 SNR \log(SNR)}}{4+\lambda_1 SNR}}_{k_2}\right)^2}\\
\label{eqn_temp3}
&\hspace{5 cm }\underbrace{e^{-\frac{\log(SNR)}{4}+\frac{\log(SNR)\lambda_1 SNR}{16+4\lambda_1 SNR}}}_{k_3}dr
\end{align}
\vspace{-0 cm}}

Let $r_0,$ $k_1,$ $k_2$ and $k_3$ be defined as shown in \eqref{eqn_temp3}. From \eqref{eqn_temp3}, the upper bound on $I$ can be written as,

{\small
\begin{align}
\nonumber
I \leq \frac{k_3}{2} \int_{r > r_0} 2(r-k_2) e^{-k_1(r-k_2)^2} dr + k_2\: k_3 \int_{r \geq r_0} e^{-k_1(r-k_2)^2} dr.
\end{align}
}

Since $\int_{r \geq r_0} e^{-k_1(r-k_2)^2} dr=\sqrt{\frac{\pi}{K}} Q \left[ (r_0-k_2)\sqrt{2k_1}\right],$ 
{
\begin{align}
\nonumber
I \leq \frac{k_3}{2 k_1} e^{-k_1(r_0-k_2)^2}+k_2\: k_3 \sqrt{\frac{\pi}{k_1}}Q \left[(r_0-k_2)\sqrt{2 k_1}\right].
\end{align}
}

 Since $Q \left[(r_0-k_2)\sqrt{2 k_1}\right] \leq 1,$ $e^{-k_1(r_0-k_2)^2}\leq 1,$ and $k_3$ can be approximated as one at high $SNR,$ substituting for $r_0,k_1,k_2$ and $k_3,$ we get,
{
\begin{align}
\nonumber
I \leq \frac{1}{2\left(1+\frac{\lambda_1 SNR}{4}\right)}+\frac{\sqrt{\pi\lambda_1 SNR \log(SNR)}}{\left(4+\lambda_1 SNR \right)\sqrt{{1}+\frac{\lambda_1 SNR}{4}}}.
\end{align}}

Since the above upper bound on $I$ falls as $SNR^{-1}$ at high SNR, $I$ has a diversity order at least 1. This completes the proof of Lemma 2.
\end{proof}
\end{lemma} 

\begin{thebibliography}{160} 
\bibitem{ZhLiLa}
S. Zhang, S. C. Liew and P. P. Lam, ``Hot topic: Physical-layer network coding,'' in \textit{Proc. ACM Annual Int. Conf. Mobile Computing and Networking}, Los Angeles, 2006, pp. 358--365.
\bibitem{PoYo_DNF}
P. Popovski and H. Yomo, ``The anti–-packets can increase the achievable throughput of a wireless multi–hop network,'' in \textit{Proc. IEEE Int. Conf. Communications}, Istanbul, 2006, pp. 3885--3890.
\bibitem{PoY}
P. Popovski and H. Yomo, ``Physical network coding in two-Way wireless relay channels,'' in \textit{Proc. IEEE Int. Conf. Communications}, Glasgow, 2007, pp. 707--712.
\bibitem{WiNaPfSp}
M. P. Wilson, K. R. Narayanan, H. D. Pfister, and A. Sprintson, ``Joint physical layer coding and network coding for bi-directional relaying,'' \textit{IEEE Trans. Info. Theory}, vol. 56, pp. 5641–-5654, Nov. 2010.
\bibitem{APT1}
T. Koike-Akino, P. Popovski and V. Tarokh, ``Optimized constellation for two-way wireless relaying with physical network coding,'' \textit{IEEE J. Sel. Areas Commun.}, vol. 27, pp. 773--787, June 2009.
\bibitem{NVR}
V. Namboodiri, V. T. Muralidharan and B. S. Rajan, ``Wireless bidirectional relaying and Latin Squares,'' in \textit{Proc. IEEE Wireless Communications and Networking Conf.}, Paris, 2012, pp. 1404--1409 (a detailed version is available in arXiv: 1110.0084v2 [cs.IT], 16 Nov. 2011).
\bibitem{VvR_ISIT}
V. T. Muralidharan and B. S. Rajan, ``Wireless network coding for MIMO two-way relaying and Latin Rectangles,'' in \textit{Proc. IEEE Int. Symp. Inf. Theory}, Cambridge, 2012. 
\bibitem{NR_Glcom}
V. Namboodiri and B. S. Rajan, ``Wirless network coding for QAM bidirectional relaying and Latin Squares,'' in \textit{Proc. IEEE Global Telecommunication Conference}, Anaheim, 2012. 
\bibitem{JaHeHuNo}
M. Janani, A. Hedayat, T. Hunter, and A. Nosratinia, ``Coded cooperation
in wireless communications: space-time transmission and iterative
decoding,'' \textit{IEEE Trans. Signal Process.}, vol. 52, pp. 362-–371,
Feb. 2004.
\bibitem{WaCaGiLa}
T. Wang, A. Cano, G. B. Giannakis and J. N. Laneman, ``High-performance cooperative demodulation
with Decode-and-Forward relays,'' \textit{IEEE Trans. Commun.}, vol. 5, pp.1427--1438, July 2007.
\bibitem{JuKi}
M. Ju and I.- M. Kim, ``ML performance analysis of the
Decode-and-Forward protocol in
cooperative diversity networks,'' \textit{IEEE Trans. Wireless Commun.}, vol. 8, pp. 3855--3867, July 2009.
\bibitem{WaGiWa}
T. Wang, G. B. Giannakis, and R. Wang, ``Smart Regenerative Relays for Link-Adaptive
Cooperative Communications,'' \textit{IEEE Trans. Commun.}, vol. 56,pp. 1950--1960, Nov. 2008.
\bibitem{WaGi}
T. Wang and G. B. Giannakis, ``Complex field network coding for multiuser
cooperative communications,'' \textit{IEEE J. Sel. Areas Commun.}, vol. 26, pp. 561--571, April 2008.
\bibitem{Rod}
Chris A. Rodger ``Recent Results on The Embedding of Latin Squares and Related Structures, Cycle Systems and Graph Designs.'', \textit{Le Matematiche}, Vol. XLVII (1992)- Fasc. II, pp. 295-311.
\bibitem{SrRa}
K. P. Srinath and B. S. Rajan, ``Low ML Decoding Complexity, Large Coding Gain, Full Rate, Full-Diversity STBCs for 2 $\times$ 2 and 4 $\times$ 2 MIMO systems,'' \textit{IEEE J. Sel. Topics Signal Process.}, vol. 3, pp. 916--927, December 2009. 
\bibitem{TaSeCa}
V. Tarokh, N. Seshadri and A. R. Calderbank, ``Space–-time codes for high data rate wireless communication: Performance criterion and code construction,'' \textit{IEEE Trans. Info. Theory}, vol. 44, pp. 744--765, March 1998.
\end{thebibliography}
\end{document}